\pdfoutput=1
\documentclass[iop]{emulateapj}
\usepackage{amsmath} 
\usepackage{aas_macros,amssymb}
\usepackage{natbib,graphicx,layouts}
\usepackage{color,hyperref}
\usepackage{bm}
\definecolor{linkcolor}{rgb}{0,0,0.5}
\definecolor{notecolor}{rgb}{0.8,0,0}
\hypersetup{colorlinks=true,linkcolor=linkcolor,citecolor=linkcolor,
  filecolor=linkcolor,urlcolor=linkcolor}

\shorttitle{Pressure smoothing in IGM}
\shortauthors{Kulkarni et al.}

\begin{document}

\title{Characterizing the Pressure Smoothing Scale of
  the Intergalactic Medium}
\author{Girish Kulkarni\altaffilmark{1,2}, 
  Joseph F.~Hennawi\altaffilmark{2}, 
  Jose O\~norbe\altaffilmark{2}, 
  Alberto Rorai\altaffilmark{1,2}, 
  Volker Springel\altaffilmark{3,4}} 
\altaffiltext{1}{Institute of Astronomy and Kavli Institute of
  Cosmology, University of Cambridge, Madingley Road, Cambridge
  CB3 0HA, UK; \email{kulkarni@ast.cam.ac.uk}}
\altaffiltext{2}{Max Planck Institute for Astronomy, K\"onigstuhl 17,
  D-69117 Heidelberg, Germany}
\altaffiltext{3}{Heidelberg Institute for Theoretical Studies,
  Schloss-Wolfsbrunnenweg 35, D-69118 Heidelberg, Germany}
\altaffiltext{4}{Zentrum f\"{u}r Astronomie der Universit\"{a}t
  Heidelberg, ARI, M\"{o}nchhofstr.\ 12-14, D-69120 Heidelberg,
  Germany}

\begin{abstract}
  The thermal state of the intergalactic medium (IGM) at $z < 6$
  constrains the nature and timing of cosmic reionization events, but
  its inference from the Ly$\alpha$ forest is degenerate with the 3-D
  structure of the IGM on $\sim 100$ kpc scales, where, analogous to
  the classical Jeans argument, the pressure of the $T\simeq
  10^{4}\,{\rm K}$ gas supports it against gravity. We simulate the
  IGM using smoothed particle hydrodynamics, and find that, at $z <
  6$, the gas density power spectrum does not exhibit the expected
  filtering scale cutoff, because dense gas in collapsed halos
  dominates the small-scale power masking pressure smoothing effects.
  We introduce a new statistic, the real-space Ly$\alpha$ flux,
  $F_\mathrm{real}$, which naturally suppresses dense gas, and is thus
  robust against the poorly understood physics of galaxy formation,
  revealing pressure smoothing in the diffuse IGM.  The
  $F_\mathrm{real}$ power spectrum is accurately described by a simple
  fitting function with cutoff at $\lambda_\mathrm{F}$, allowing us to
  rigorously quantify the pressure smoothing scale for the first time:
  we find $\lambda_\mathrm{F} = 79$ kpc (comoving) at $z=3$ for our
  fiducial thermal model. This statistic has the added advantage that
  it directly relates to observations of correlated Ly$\alpha$ forest
  absorption in close quasar pairs, recently proposed as a method to
  measure the pressure smoothing scale. Our results enable one to
  quantify the pressure smoothing scale in simulations, and ask
  meaningful questions about its dependence on reionization and
  thermal history. Accordingly, the standard description of the IGM in
  terms of the amplitude $T_0$ and slope $\gamma$ of the
  temperature-density relation $T = T_0(\rho\slash
  {\bar \rho})^{\gamma-1}$ should be augmented with a third pressure
  smoothing scale parameter $\lambda_\mathrm{F}$.
\end{abstract}

\keywords{dark ages, reionization, first stars -- galaxies: high-redshift --  intergalactic medium -- large-scale structure of the universe -- quasars: absorption lines}

\section{Introduction}
\label{sec:intro}

The intergalactic medium (IGM) is a rich repository of cosmic history.
It contains most of the mass of the universe and records a variety of
baryonic and non-baryonic processes that occur as the universe evolves
\citep{2009RvMP...81.1405M}.  This makes the IGM a valuable
cosmological laboratory for testing models of structure formation.
The most readily observable probe of the IGM is the Ly$\alpha$ forest,
the collection of absorption lines seen at redshifts $z\sim 2$--$5$ in
spectra of high-redshift quasars.  Hydrodynamical simulations of
structure formation in a $\Lambda$CDM universe show that the
Ly$\alpha$ forest results from the interaction between a background of
UV photons created by active galactic nuclei (AGN) and star-forming
galaxies, and fluctuations in gas density sourced by gravitational
instability \citep{1994ApJ...437L...9C, 1996ApJ...457L..51H,
  1997ApJ...485..496Z, 2000ApJ...543....1M, 2001MNRAS.327..296M,
  2002ApJ...580..634C, 2004MNRAS.354..684V, 2015MNRAS.446.3697L}.
This picture successfully explains several statistical properties of
the Ly$\alpha$ forest, such as the H~\textsc{i} column density
distribution, and the line-of-sight power spectrum and probability
density function (PDF) of the transmitted flux.

The Ly$\alpha$ forest can be used to probe the evolution of the
thermal state of the IGM at redshifts $z\sim 2$--$5$.  On cosmological
time scales, the thermal state of the IGM is expected to evolve
through four stages
\citep[e.g.,][]{2001PhR...349..125B,2012RPPh...75h6901P}: (1) Between
recombination ($z\sim 1100$) and $z=147$, the cosmic gas temperature
is coupled to the cosmic microwave background (CMB) temperature via
Compton scattering off a small fraction of residual free electrons.
As a result the gas temperature falls as $T_\mathrm{gas}\propto (1+z)$
from $T_\mathrm{gas}\gtrsim 10^3$~K at recombination to
$T_\mathrm{gas}\sim 400$~K at $z=147$. (2) At lower redshifts
($z<147$), the free electron fraction is no longer sufficient to
couple gas and CMB temperatures, and gas cools adiabatically as
$T_\mathrm{gas}\propto (1+z)^2$ down to $\mathbf{T_\mathrm{gas}\sim
2}$~K at $z=10$. (3) As galaxies start forming at $z=10$, UV radiation
from young, massive stars and possibly a population of faint AGN
deposits large amount of energy in the IGM on time scales of $\sim
500$~Myr.  In this relatively short period of time---the ``epoch of
hydrogen reionization''---most of the H~\textsc{i} and He~\textsc{i}
content of the IGM is ionized to H~\textsc{ii} and He~\textsc{ii}
respectively.  This also heats the IGM via photoionization heating so
that by $z\sim 6$ the IGM temperature is $T_\mathrm{gas}\sim
10^4$~K \citep{2010MNRAS.406..612B, 2012MNRAS.419.2880B}.  (4) Due to
the small neutral hydrogen fraction ($x_\mathrm{HI}\sim 10^{-4}$) in
the IGM at the end of the epoch of hydrogen reionization, UV photons
can propagate through the IGM to large distances without absorption.
As a result, a UV background is established.  The IGM remains in
ionization equilibrium with this background and begins to cool as the
Universe expands.  However, AGN activity peaks at $z\sim 3$, resulting
in an abundance of hard photons that ionize He~\textsc{ii} to
He~\textsc{iii} and likely reheats the IGM to higher temperatures.
This ``epoch of helium reionization'' probably ends at $z\sim 2.7$
\citep{2010ApJ...722.1312S, 2011ApJ...733L..24W, 2013ApJ...765..119S,
  2014arXiv1405.7405W} after which the IGM is again expected to cool
down to the present epoch.  By studying the Ly$\alpha$ forest one can
constrain the thermal state of the IGM providing insights into the
Universe's thermal history.

It is well known that the balance between cooling due to Hubble
expansion and heating due to the gravitational collapse and
photoionization heating give rise to a well-defined
temperature-density relationship in the IGM
\citep{1997MNRAS.292...27H}\footnote{The temperature-density relation
  is often also called the effective equation of state of the IGM.
  The thermodynamic equation of state is that of an ideal gas.}, which
can be written as
\begin{equation}
  T=T_0\left(\frac{\rho}{\bar\rho}\right)^{\gamma-1},
  \label{eqn:trho_relation}
\end{equation}
where $T_0$ is the temperature at the mean density $\bar\rho$.
Immediately after the reionization of H~\textsc{i} ($z<6$) or
He~\textsc{ii} ($z<3$), $T_0$ is likely to be around $2\times 10^4$~K
and $\gamma\sim 1$ \citep{1997MNRAS.292...27H}.  At lower redshifts,
$T_0$ decreases adiabatically while $\gamma$ is expected to increase
and asymptotically approach a value of $1.62$
\citep{1997MNRAS.292...27H}.  The remarkable fact that the mean
temperature-density relation of the baryons in the IGM can be well
approximated by a simple power law is behind the standard
observational approach of measuring the thermal state of the IGM from
the Ly$\alpha$ forest: statistical properties of the forest are
measured and compared to hydrodynamical simulations to constrain $T_0$
and $\gamma$.  Statistics considered for this purpose are
line-of-sight power spectrum of transmitted flux
\citep{2001ApJ...557..519Z, 2009MNRAS.399L..39V}, PDF of wavelet
amplitudes in a wavelet decomposition of the forest
\citep{2000MNRAS.317..989T, 2002MNRAS.332..367T, 2010ApJ...718..199L,
  2012MNRAS.424.1723G}, the average local curvature of the spectrum
\citep{2011MNRAS.410.1096B, 2014MNRAS.441.1916B}, PDF of transmitted
flux values \citep{2001ApJ...562...52M, 2007MNRAS.382.1657K,
  2008MNRAS.386.1131B, 2012MNRAS.422.3019C, 2012MNRAS.424.1723G,
  2015ApJ...799..196L}, and the slope of the $b$-parameter
distribution \citep{1998MNRAS.298L..21H, 2000MNRAS.315..600T,
  2000ApJ...534...41R, 2000ApJ...534...57B, 2000MNRAS.318..817S,
  2001ApJ...562...52M, 2002ApJ...567L.103T, 2012ApJ...757L..30R,
  2014MNRAS.438.2499B}.

However, these measurements rely on the longitudinal, one-dimensional,
structure of the Ly$\alpha$ forest, which is also influenced by the
three-dimensional pressure smoothing.  At large scales and low
densities, baryons in the IGM are expected to follow the underlying
dark matter distribution.  However, at small scales and high
densities, baryons experience pressure forces that prevent them from
tracing the collisionless dark matter.  This pressure results in an
effective three-dimensional smoothing of the baryon distribution
relative to the dark matter, at a characteristic scale known as the
pressure smoothing scale.  This pressure smoothing affects
absorption lines in the Ly$\alpha$ forest by reducing power in the
distribution of H~\textsc{i} at scales smaller than the pressure
smoothing scale. This effect, which reduces the small-scale power in
the longitudinal transmitted flux power spectrum, is degenerate with a
similar but one-dimensional reduction due to thermal
broadening \citep{2010MNRAS.404.1281P,
2010MNRAS.404.1295P,2013ApJ...775...81R, 2015arXiv150205715G}.

This degeneracy has been largely ignored by previous measurements of
the IGM temperature, all of which use line-of-sight information from
the Ly$\alpha$ forest (but see \citet{2011MNRAS.410.1096B} and
\citet{2014arXiv1410.1531P}).  As a result, published constraints on
$T_0$ and $\gamma$ are confusing and sometimes contradictory.  For
example, measurements of $\gamma$ do not agree with each other; some
measurements even suggest an inverted temperature-density relation
($\gamma<1$), which is difficult to obtain in canonical IGM models
\citep{2008MNRAS.386.1131B, 2009MNRAS.399L..39V, 2012MNRAS.422.3019C,
  2012MNRAS.424.1723G}.\footnote{This can however be achieved in more
  exotic models such as TeV blazar heating \citep{2012ApJ...752...23C,
    2012MNRAS.423..149P}.} Consequently, an important goal of IGM
studies is to break this degeneracy and bring robustness to
measurements of the thermal state of the IGM.

\begin{figure*}
  \begin{center}
  \begin{tabular}{cc}
    \includegraphics*[width=\columnwidth]{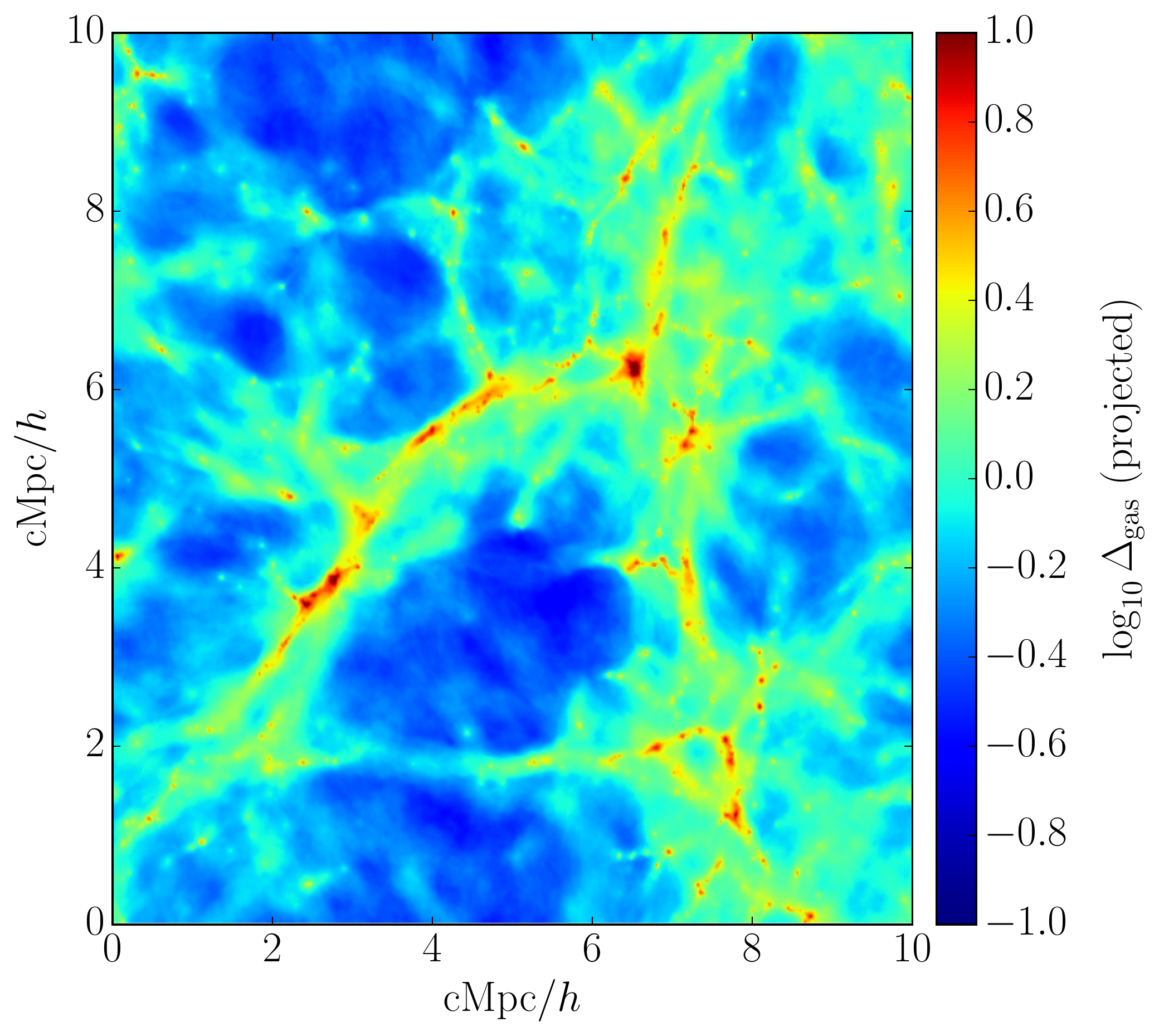} &
    % gas: gas_slice.py 
    \includegraphics*[width=\columnwidth]{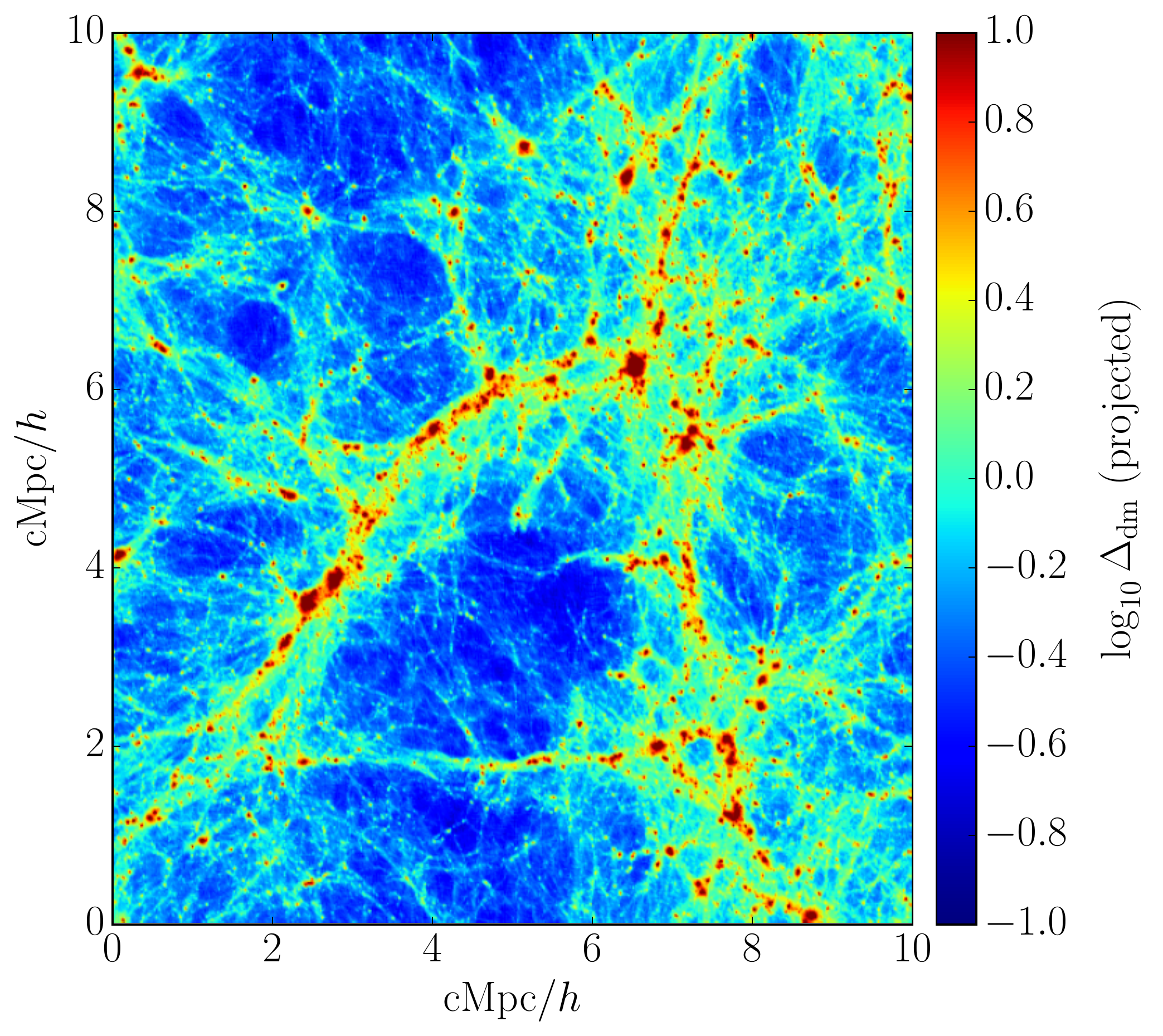} \\
    % dm: dm_slice.py 
  \end{tabular}
  \end{center}
  \caption{Projected density distributions of gas (left) and dark
    matter (right) at $z=3$ in our fiducial simulation, showing
    pressure smoothing of gas relative to dark matter.  The density at
    each point is an average for a column approximately 5 Mpc$/h$
    long.}
  \label{fig:gas_and_dm_slices}
\end{figure*}

The degeneracy can be broken by a suitable transverse measurement that
is sensitive only to pressure smoothing.  \citet{2013ApJ...775...81R}
showed that this can be achieved by measuring coherence of Ly$\alpha$
forest absorption in close quasar pairs.  Spectra of close quasar
pairs will become increasingly coherent as the transverse separation
between the quasars approaches the pressure smoothing
scale. \citet{2013ApJ...775...81R} demonstrated that this effect can
be exploited by measuring the probability distribution function of
phase angle differences between homologous line-of-sight Fourier modes
in quasar pair spectra. This new technique has tremendous promise for
direct measurements of the pressure smoothing scale, and
\citet{2013ApJ...775...81R} showed that applying it to a realistic
data set should yield measurements with 5\% precision.

The precision with which quasar pair measurements will be able to
characterize the pressure smoothing scale motivates us to better understand
pressure smoothing in the IGM.  Classically, a simple characterization
of the pressure smoothing scale is given by \citep{2008gady.book.....B}
\begin{equation}
  \lambda_\mathrm{F,Jeans}^2(t) = \frac{c_s^2(t)a(t)}{4\pi G\bar\rho_\mathrm{m0}},
  \label{eqn:jeans}
\end{equation}
in comoving units, where $\bar\rho_\mathrm{m0}$ is the average
comoving matter density, $a$ is the cosmological scale factor, and
$c_s$ is the sound speed.  Thus the pressure smoothing scale is a function of
the instantaneous gas temperature.  Nevertheless, in an expanding
universe with an evolving thermal state, the pressure smoothing scale at a
given epoch is expected to depend on the entire thermal history,
because fluctuations at earlier times expand or fail to collapse
depending on the IGM temperature at that epoch.  In the limit of
linear density perturbations, an analytical estimate of the dependence
of the pressure smoothing scale on thermal history can be obtained
\citep{1998MNRAS.296...44G}.  The resultant pressure smoothing scale, termed filtering scale, can be
written as
\begin{equation}
\begin{split}
\lambda_\mathrm{F,GH98}^2(t)=\frac{4\pi G\bar\rho_\mathrm{m0}}{b(t)}
&\int_0^tdt^\prime\frac{b(t^\prime)
  \lambda_\mathrm{F,Jeans}^2(t^\prime)}{a(t^\prime)}\\&\times\int_{t^\prime}^t\frac{dt^{\prime\prime}}{a^2(t^{\prime\prime})},
\end{split}
\label{eqn:gh98}
\end{equation}
in comoving units, where $\bar\rho_\mathrm{m0}$ is the average
comoving matter density, $b$ is the growth factor of linear
perturbations, and $\lambda_\mathrm{F,Jeans}$ is given by
Equation~(\ref{eqn:jeans}).  Note that the filtering scale
$\lambda_\mathrm{F,GH98}$ is larger than $\lambda_\mathrm{F,Jeans}$
before reionization, and smaller than $\lambda_\mathrm{F,Jeans}$
(typically by about a factor of 2) after reionization.  At redshifts
probed by the Ly$\alpha$ forest, typical values of
$\lambda_\mathrm{F,GH98}$, assuming simple models for the IGM thermal
state, are about a hundred comoving kiloparsecs.  With this filtering
scale in hand, it is usually assumed that baryonic density is given by
the dark matter density smoothed by a quadratic
\citep{1997ApJ...479..523B} or a Gaussian kernel
\citep{2003ApJ...583..525G, 2013ApJ...775...81R} with width
$\lambda_\mathrm{F,GH98}$.

However, this description of the pressure smoothing in the IGM suffers
from two problems.  First, scales comparable to the pressure smoothing
scale, i.e., scales of order hundreds of comoving kpc, are highly
nonlinear at redshifts $z\sim 2$--$5$ probed by the Ly$\alpha$ forest.
Therefore, the above linear perturbation theory description of
$\lambda_\mathrm{F}$ breaks down at these redshifts.  This means that
the dependence of the IGM pressure smoothing scale on the thermal
history is likely more complicated than Equation~(\ref{eqn:gh98}).
Secondly, it is unclear that pressure smoothing of the baryons can be
well described as Gaussian or quadratic smoothing of underlying dark
matter density, because at scales of $\sim 100\,$ comoving
kiloparsecs, galaxy formation physics could significantly influence
the baryon distribution.  Both of these limitations will affect the
interpretation of the pressure smoothing scale from observations of
the Ly$\alpha$ forest.  The goal of this paper is to shed light on the
pressure smoothing scale of the IGM using high-resolution
hydrodynamical simulations.  We ask if the pressure smoothing in the
IGM is well described by the approximations in
Equations~(\ref{eqn:jeans}) and (\ref{eqn:gh98}), and present an
improved characterization of the pressure smoothing scale.

The structure of this paper is as follows: In Section~\ref{sec:past}
we review the quadratic and Gaussian descriptions of the IGM and the
motivations behind them.  We present our simulations in
Section~\ref{sec:sims}.  We analyze our numerical results and present
a new characterization of the IGM pressure smoothing scale in
Section~\ref{sec:results}.  Section~\ref{sec:conc} contains a summary
of our main conclusions.  We work with a $\Lambda$CDM cosmological
model with $\Omega_\mathrm{b}=0.045$, $\Omega_\mathrm{m}=0.26$,
$\Omega_\Lambda=0.73$, $h=0.71$, $n=0.96$, $\sigma_8=0.80$
\citep{2013ApJS..208...19H}.  All distances are in comoving units.

\begin{figure*}
  \begin{center}
  \begin{tabular}{cc}
    \includegraphics*[scale=0.45]{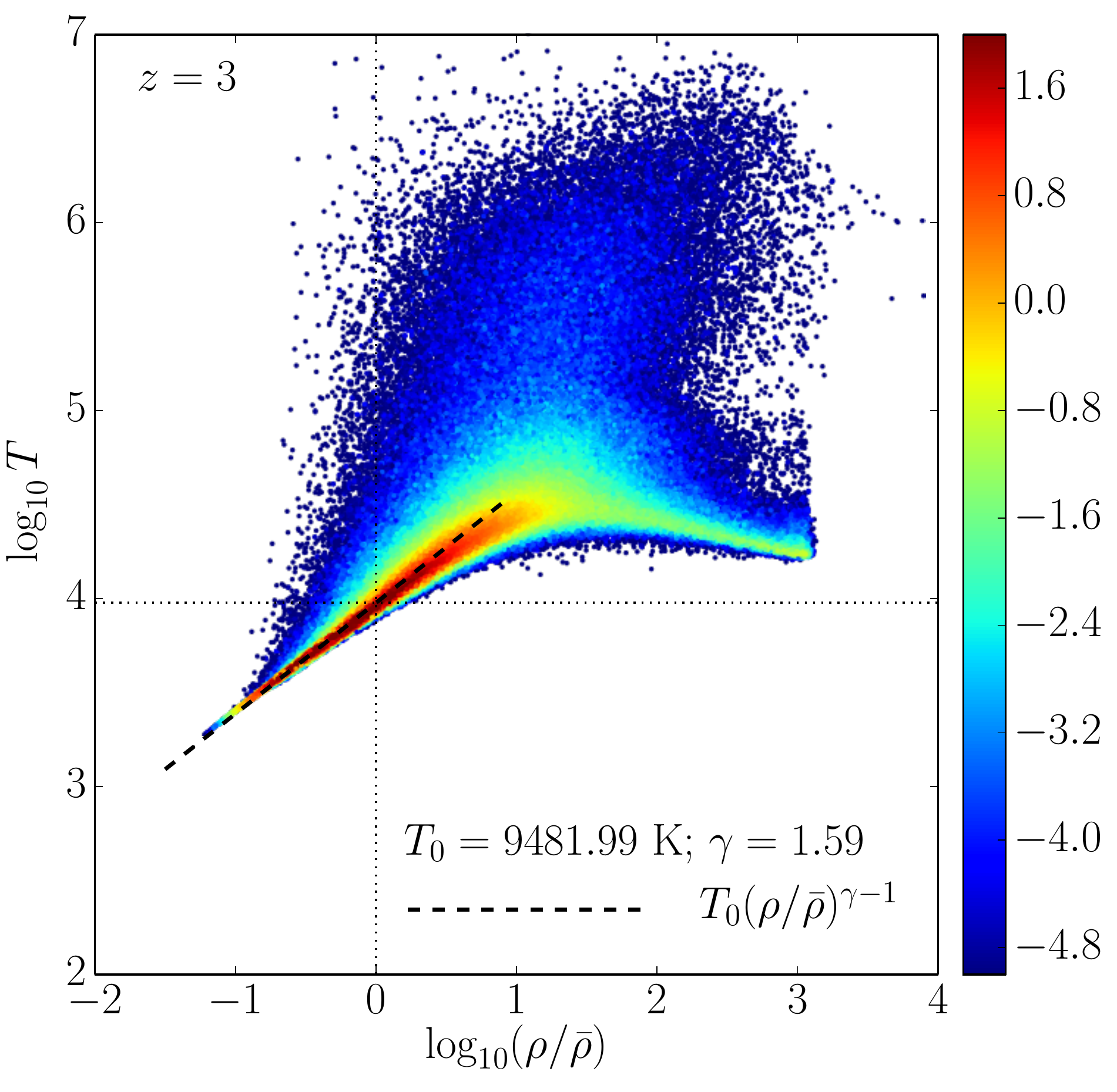}
    &
    \includegraphics*[scale=0.45]{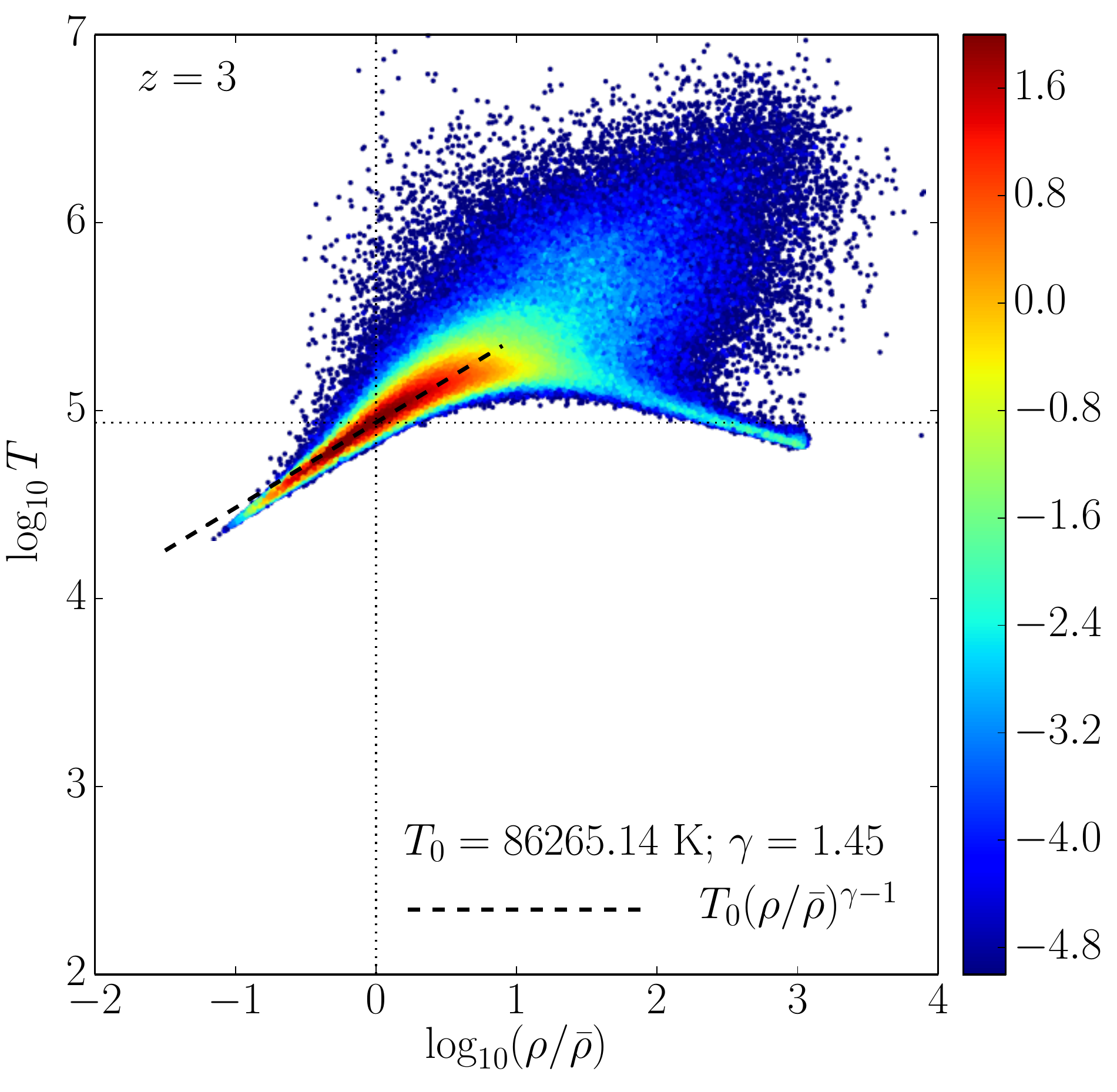}
    \\
  \end{tabular}
  \end{center}
  \caption{Comparison of the phase-space distributions of gas
    particles in two of the simulations presented in this paper.
    (Left panel describes the fiducial simulation.)  Color scale
    describes density of points on the plot; most particles are on the
    temperature-density relation shown by the dashed black line.}
  \label{fig:trho}
\end{figure*}

\section{Analytical Estimates}
\label{sec:past}

In this section, we review the linear theory results that are
traditionally used to describe the pressure smoothing scale of the
IGM.  The IGM is conventionally described as an ideal gas with a
polytropic equation of state, $p=K\rho^{5/3}$.  Then, in comoving
coordinates, linear evolution of density perturbations is given by
\begin{align}
  &\frac{d^2\delta_\mathrm{b}}{dt^2} + 2H\frac{d\delta_\mathrm{b}}{dt}
  = 4\pi G(\bar\rho_\mathrm{b}\delta_\mathrm{b} +
  \bar\rho_\mathrm{m}\delta_\mathrm{m}) -
  \frac{c_s^2}{a^2}k^2\delta_\mathrm{b}, \nonumber
  \\ &\frac{d^2\delta_\mathrm{m}}{dt^2} +
  2H\frac{d\delta_\mathrm{m}}{dt} = 4\pi
  G(\bar\rho_\mathrm{b}\delta_\mathrm{b} +
  \bar\rho_\mathrm{m}\delta_\mathrm{m}),
\label{eqn:linevol}
\end{align}
where $\delta_\mathrm{b}(t,\bm{k})$ and $\delta_\mathrm{m}(t,\bm{k})$
are Fourier amplitudes of baryon and dark matter density contrasts
($\delta\equiv\rho/\bar\rho-1$) respectively and $\bar\rho_\mathrm{b},
\bar\rho_\mathrm{m}\propto a^{-3}$.  The ratio
$\delta_\mathrm{b}(t,\bm{k})/\delta_\mathrm{m}(t,\bm{k})$ will reveal
the pressure smoothing scale $k_\mathrm{F}$ at which pressure support equals
gravitational force.

\subsection{Linear evolution for an adiabatic thermal history}

We can ignore $\bar\rho_\mathrm{b}$ relative to $\bar\rho_\mathrm{m}$
in the gravitational source term on the right hand sides of
Equations~(\ref{eqn:linevol}).  The dark matter evolution can then be
solved as usual to obtain the growing mode $b(z)$
\citep{1980lssu.book.....P}.  For baryonic perturbations, a general
solution can be obtained at $z>2$ where $b\sim a$, if we assume that
the temperature at the mean density, $T_0$ evolves as $1/a$
\citep{1992A&A...266....1B}.  This solution can be written as
\begin{equation}
  \frac{\delta_\mathrm{b}(t,\bm{k})}{\delta_\mathrm{m}(t,\bm{k})} =
  \frac{1}{1+\lambda_\mathrm{J}^2k^2},
  \label{eqn:bidavid}
\end{equation}
where we have only retained the growing mode.  The parameter
$\lambda_\mathrm{J}$ is defined as
\begin{equation}
  \lambda_\mathrm{J}^2 = \frac{c_s^2}{4\pi G\bar\rho_\mathrm{m}a^2},
\label{eqn:xb}
\end{equation}
which is just the classical Jeans scale of Equation~(\ref{eqn:jeans}).
Due to Equation~(\ref{eqn:xb}), for a polytropic IGM, the assumption
that $T_0\propto 1/a$ results in a constant $\lambda_\mathrm{J}$ at
all times.  Equation~(\ref{eqn:bidavid}) describes the behavior of
baryons relative to the dark matter to linear order in
$\delta_\mathrm{b}$, at redshifts $z>2$, under the assumption that
$T_0\propto 1/a$.  It shows that, in linear theory, baryonic density
fluctuations are damped relative to the dark matter fluctuations by a
quadratic filter $1+\lambda_\mathrm{J}^2k^2$.  At large scales
($k\ll\lambda_\mathrm{J}$), baryons follow the dark matter
($\delta_\mathrm{b}=\delta_\mathrm{m}$) but at small scales
($k\gg\lambda_\mathrm{J}$), baryons are smoothed out by pressure
support
($\delta_\mathrm{b}=\delta_\mathrm{m}/\lambda_\mathrm{J}^2k^2$).\footnote{A
  somewhat more general solution can be obtained for
  Equations~(\ref{eqn:linevol}) by assuming $T_0\propto a^{-\alpha}$,
  which corresponds to a power law evolution of $\lambda_\mathrm{J}$
  \citep{1992A&A...266....1B}.  (The above analysis corresponds to
  $\alpha=1$).  However, this solution has the same asymptotic
  behavior as Equation~(\ref{eqn:bidavid}).  Thus the quadratic
  filter in Equation~(\ref{eqn:bidavid}) is a general feature of power
  law adiabatic thermal histories in the linear approximation.}

\subsection{Linear evolution for an arbitrary thermal history}

However, as discussed in Section~\ref{sec:intro} above, in general the
thermal history of the IGM is expected to be different from $T\propto
1/a$.  For example, $T$ is expected to increase during cosmic
reionization events, and additionally photoionization heating due to
the UV background will generally result in different temperature
evolution. In general, for an arbitrary thermal history $T(a)$ a
generalization of Equation~(\ref{eqn:bidavid}) cannot be obtained
analytically.  But an approximate result can be derived to first order
in $k^2$, i.e., at large scales \citep{1998MNRAS.296...44G}.  This
solution can be written as
\begin{equation}
  \frac{\delta_\mathrm{b}(t,\bm{k})}{\delta_\mathrm{m}(t,\bm{k})} = 1-k^2\lambda_\mathrm{F}^2, 
\end{equation}
where 
\begin{equation}
\lambda_\mathrm{F}^2=\frac{4\pi G\bar\rho_\mathrm{m0}}{b(t)}
\int_0^tdt^\prime\frac{b(t^\prime)\lambda_\mathrm{J}^2(t^\prime)}{a(t^\prime)}\int_{t^\prime}^t\frac{dt^{\prime\prime}}{a^2(t^{\prime\prime})},
\label{eqn:kf}
\end{equation}
in which the thermal history enters via $\lambda_\mathrm{J}$ and
$\bar\rho_\mathrm{m0}$ is the average matter density at $z=0$ (recall
that $\lambda_\mathrm{J}$ is defined in Equation~(\ref{eqn:xb})). Note
that this describes the linear theory evolution of the IGM for an
arbitrary thermal history only at large scales
($k^2\lambda_\mathrm{F}^2\ll 1$).  A closed-form solution of
Equations~(\ref{eqn:linevol}) to all orders of
$k^2\lambda_\mathrm{F}^2$ is not known.  However, by solving
Equations~(\ref{eqn:linevol}) numerically, \citet{2003ApJ...583..525G}
show that the full solution is described to a good accuracy by a
Gaussian of the form
\begin{equation}
  \frac{\delta_\mathrm{b}(t,\bm{k})}{\delta_\mathrm{m}(t,\bm{k})}=\exp\left(-k^2\lambda_\mathrm{F}^2\right),
  \label{eqn:fullsoln}
\end{equation}
where $\lambda_\mathrm{F}$ is defined in Equation~(\ref{eqn:kf}).
Note that this solution describes the density fields only in the limit
of linear perturbations.  Typically, Equation~(\ref{eqn:fullsoln})
predicts a pressure smoothing scale of $\lambda_\mathrm{F}\sim 100$~kpc
\citep{1998MNRAS.296...44G}.

\subsection{Nonlinear evolution}

Equations~(\ref{eqn:bidavid}) and (\ref{eqn:fullsoln}) describe
pressure smoothing of the baryons in the linear perturbation theory
regime, i.e., in the limit of small density ($\delta<1$).  The
generalization of these results to the nonlinear limit is unknown.  By
arguing that the Ly$\alpha$ forest is only sensitive to moderate
overdensities ($\Delta\lesssim 10$), Equations~(\ref{eqn:bidavid}) and
(\ref{eqn:fullsoln}) have been used in semi-analytical descriptions of
the forest \citep{1995ApJ...449..476R, 1997ApJ...479..523B,
  2001MNRAS.322..561C, 2001ApJ...559...29C}.  Several authors have
also used these results to simulate the IGM using so-called
pseudo-hydrodynamical methods \citep{1995A&A...295L...9P,
  1998MNRAS.296...44G, 1998ApJ...495...44C}.  In terms of their
predictions for simple cumulative statistics of absorption lines in
the Ly$\alpha$ forest, these methods give results that are close to
within 10\% of hydrodynamical simulations \citep{2001MNRAS.324..141M}.

Nonetheless, the application of Equations~(\ref{eqn:bidavid}) and
(\ref{eqn:fullsoln}) to characterize pressure smoothing at the scales
and redshifts probed by the Ly$\alpha$ forest is problematic.
Equations~(\ref{eqn:bidavid}) and (\ref{eqn:fullsoln}) typically
predict pressure smoothing scales of order 100~kpc.  At
redshifts $z = 2$--$5$, such small scales are highly nonlinear, as the
variance in the density distribution at these scales is dominated by
collapsed structures.  Baryons at these densities are affected by the
physics of galaxy formation, as well as gas outflows and inflows.
Therefore, in order to understand pressure smoothing in the IGM, it
would be necessary to somehow distinguish, at scales of $\sim 100$
kpc, the low density IGM that is likely unaffected by galaxy formation
from the high density IGM that is affected by it.  But where to draw
the line between these two density regimes is not clear.  Also, unlike
in the linear regime ($\Delta< 1$), the pressure smoothing scale
$\lambda_\mathrm{F}$ is expected to be dependent on the local density
in the quasi-linear regime ($\Delta\lesssim 10$) probed by the
Ly$\alpha$ forest, but this dependence is unknown, as the classical
result of Equation~(\ref{eqn:xb}) is not guaranteed to hold at
densities of order $1$--$10$.  Therefore, even if the low density IGM
is somehow isolated, we are still left with the task of characterizing
its pressure smoothing.

In this paper we address these two issues, and present a method of
isolating the low-density IGM in a way which enables characterization
of the pressure smoothing scale that not only describes the pressure smoothing
in the IGM, but also provides an important tool for interpreting
pressure smoothing scale measurements from the Ly$\alpha$ forest.

\section{Hydrodynamical Simulations}
\label{sec:sims}

To understand pressure smoothing in the IGM we run high resolution
cosmological hydrodynamical simulations using the energy- and
entropy-conserving TreePM smoothed particle hydrodynamical (SPH) code
\textsc{p-gadget-3}, which is an updated version of the
\textsc{gadget} code \citep{2001NewA....6...79S, 2005MNRAS.364.1105S}.
In addition to the cosmological evolution of baryons and dark matter,
this code implements photoionization and photoheating of baryons by
calculating the equilibrium ionization balance of hydrogen and helium
in a optically thin UV background, which is taken from an updated
version of the model of \citet{1996ApJ...461...20H} presented by
\citet{1999ApJ...511..521D}. Radiative cooling is implemented by
taking into account cooling via two-body processes such as collisional
excitation of H~\textsc{i} and He~\textsc{ii}, collisional ionization
of H~\textsc{i}, He~\textsc{i}, and He~\textsc{ii}, recombination, and
Bremsstrahlung \citep{1996ApJS..105...19K}.  Likewise,
\textsc{p-gadget-3} also includes inverse Compton cooling off the CMB
\citep{1986ApJ...301..522I}, which can be an important source of
cooling at high redshifts.  For temperatures in the range
$10^4$--$10^5$~K, at redshifts probed by the Ly$\alpha$ forest,
recombination cooling is the dominant cooling mechanism (we ignore
metal enrichment and its effect on cooling rates, which is a good
approximation for the IGM).

Baryons with overdensities higher than a few hundred times the
background density are expected to form galaxies.  But as these high
density regions are not the subject of our study, we simplify their
treatment.  In the simulations presented here, all gas particles with
temperature less than $10^5$~K and overdensity of more than a thousand
times the mean baryon density are converted to collisionless stars and
removed from the hydrodynamical calculation
\citep{2004MNRAS.354..684V}.  This speeds up the simulations, while
leaving the low-density IGM unaffected.  We use the {\tt
  QUICK\_LYALPHA} flag in \textsc{p-gadget-3} for this purpose, and in
what follows we will also vary the value of this threshold to
understand its impact on our results.

\begin{figure}
  \begin{center}
    \includegraphics*[width=\columnwidth]{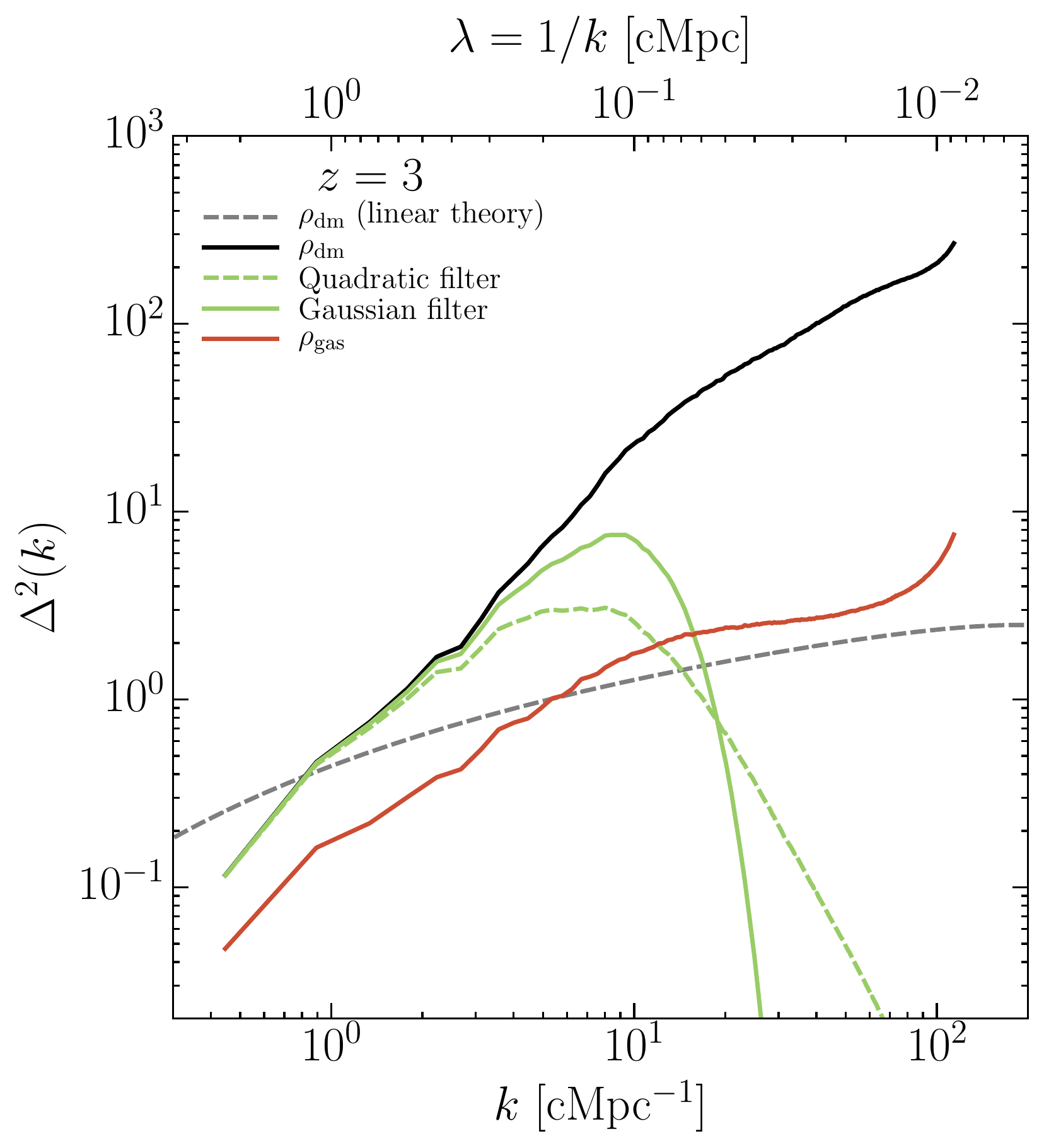}
    % ps.py (case 16) 
  \end{center}
  \caption{Gas (red curve) and dark matter (black curve) density power
    spectra at $z=3$ in our fiducial simulation.  Also shown are the
    linearly extrapolated matter power spectrum (dashed grey curve),
    and the quadratic (dashed green curve) and Gaussian (solid green
    curve) filters from Equations~(\ref{eqn:bidavid}) and
    (\ref{eqn:fullsoln}).  The gas density power spectrum does not
    show a cut-off.}
  \label{fig:ps}
\end{figure}

Simulations presented in this paper were all done in a cubic box of
length $10$ comoving $h^{-1}$Mpc on the side.  Periodic boundary
conditions were imposed.  The number of dark matter and gas elements
are both initially $512^3$.  This corresponds to a dark matter
particle mass of $M_\mathrm{dm}=4.5\times 10^5$ $h^{-1}$M$_\odot$ and
gas particle mass of $M_\mathrm{gas}=9.3\times 10^4$
$h^{-1}$M$_\odot$.  This gives us a mean inter-particle separation of
better than 20 comoving $h^{-1}$kpc, which sets our spatial resolution
in the lowest density regions of the IGM.  This resolution is
sufficient for simulating the Ly$\alpha$ forest
\citep{2009MNRAS.398L..26B,2015MNRAS.446.3697L}.  Specifically, at
$z=2$ it has been found that a mass resolution of
$M_\mathrm{gas}=1.6\times 10^6$ $h^{-1}$M$_\odot$ is sufficient for
convergence in the mean Ly$\alpha$ flux and and flux power spectrum in
\textsc{gadget} \citep{2009MNRAS.398L..26B}.  At higher redshifts, the
mass resolution requirement grows stronger: at $z=5$ a mass resolution
of about $M_\mathrm{gas}=2\times 10^5$ $h^{-1}$M$_\odot$ is required.
Our mass resolution meets both these requirements \citep[see
  also][]{1999ApJ...517...13B}.  However, in their study
\citet{2009MNRAS.398L..26B} also found that a box size of $10$
comoving $h^{-1}$Mpc is too small for convergence in these quantities,
giving errors of about 10\% for the flux power spectrum \citep[see
  also][]{2015MNRAS.446.3697L}.  They find that a box size of at least
$20$ comoving $h^{-1}$Mpc was required at $z=2$.  At $z=5$, a box of
size $40$ comoving $h^{-1}$Mpc was required.  When the box size is
small, the error due to the missing large-scale modes and the coupling
of density perturbation modes in the box to modes larger than the box
size is large.  This results in the loss of the overall amplitude of
the power spectrum of density perturbations \citep[see
  also][]{2004MNRAS.350.1107M, 2005ApJ...635..761M,
  2006MNRAS.365..231V, 2009MNRAS.393..723T, 2010ApJ...718..199L,
  2015MNRAS.446.3697L}. However, as we will see below, this overall
loss of power is not a concern for the purpose of characterizing the
pressure smoothing scale.

Figure \ref{fig:gas_and_dm_slices} shows distributions of gas and dark
matter density from the fiducial simulation at $z=3$ in slices
approximately 20 comoving kpc$/h$ thick.  Gas and dark matter
distributions are similar on large scales.  But on small scales, the
gas distribution is more diffuse.  This difference is more prominent
in quasi-linear structure such as filaments and is due to the pressure
smoothing that we aim to characterize in this paper. 

\begin{figure}
  \begin{center}
    \includegraphics*[width=\columnwidth]{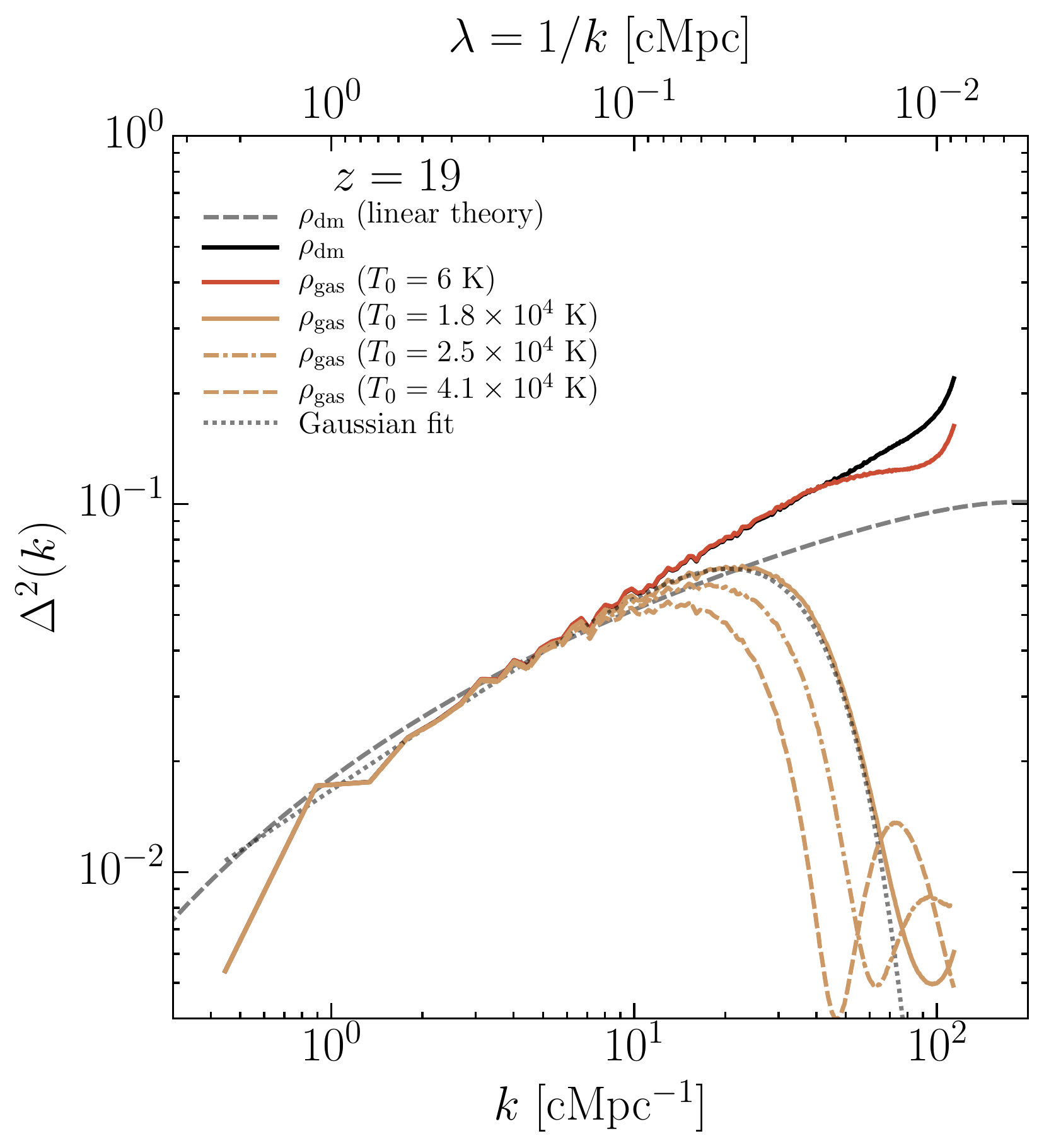}
    % ps.py case 11 
  \end{center}
  \caption{Gas density power spectrum at $z=19$.  The solid red curve
    shows the power spectrum in our fiducial run; solid, dashed, and
    dot-dashed beige curves show power spectra in simulations with
    high $T_0$ at this redshift.  The dotted curve is a Gaussian fit
    to one of the power spectra.  Also shown are the linear and
    nonlinear dark matter power spectra.  In contrast to Figure
    \ref{fig:ps}, the gas power spectrum shows a Gaussian cut-off in
    accordance with Equation~(\ref{eqn:fullsoln}) at this high
    redshift when most scales in the box are linear.  Note that in the
    fiducial simulation the pressure smoothing scale is too low to be resolved
    due to the low temperature at this high redshift.}
  \label{fig:ps_z19_multipleT}
\end{figure}

Apart from the fiducial simulation, we also present results from four
other simulations, which differ from the fiducial simulation only in
their thermal evolution.  In these simulations, the photoheating rate
of the H~\textsc{i}, He~\textsc{i}, and He~\textsc{ii} in the IGM is
enhanced artificially by constant factors of 2, 5, 10, and 50
\citep{2011MNRAS.410.1096B} resulting in different thermal histories.
The temperature of the IGM at mean density, $T_0$ at $z=3$ for these
simulations is, $1.4\times 10^4$~K, $2.4\times 10^4$~K, $3.6\times
10^4$~K, and $8.6\times 10^4$~K respectively. In the fiducial
simulation $T_0$ at this redshift is $9.5\times 10^3$~K.  Figure
\ref{fig:trho} shows the phase space distribution of gas particles at
$z=3$ in the fiducial simulation and in the high-temperature
simulation with $T_0=8.6\times 10^4$~K.  (Photoheating rate in this
simulation is enhanced by a factor of 50 relative to the fiducial
rate.)  Both distributions have familiar features: an overwhelming
fraction of low density ($\Delta<10$) gas particles reside on the
power law temperature-density relationship of the form given by
Equation~(\ref{eqn:trho_relation}) shown by the dashed line.  A
fraction of the highest density gas ($\Delta>100$) has cooled in
collapsed haloes, and the rest is shock heated to virial temperatures
exceeding $10^6$~K.  At this high redshift ($z=3$) the low-density
high-temperature warm hot intergalactic medium (WHIM;
\citealt{1999ApJ...514....1C, 2006ApJ...650..560C,
  2001ApJ...552..473D}) is still sparsely populated.  The phase space
diagrams in Figure \ref{fig:trho} show a cut-off at $\Delta=10^3$ for
temperatures below $10^5$~K, which is due to the simple star formation
recipe described above, which converts these cool high-density gas
particles to stars.

\begin{figure}
  \begin{center}
    \includegraphics*[width=\columnwidth]{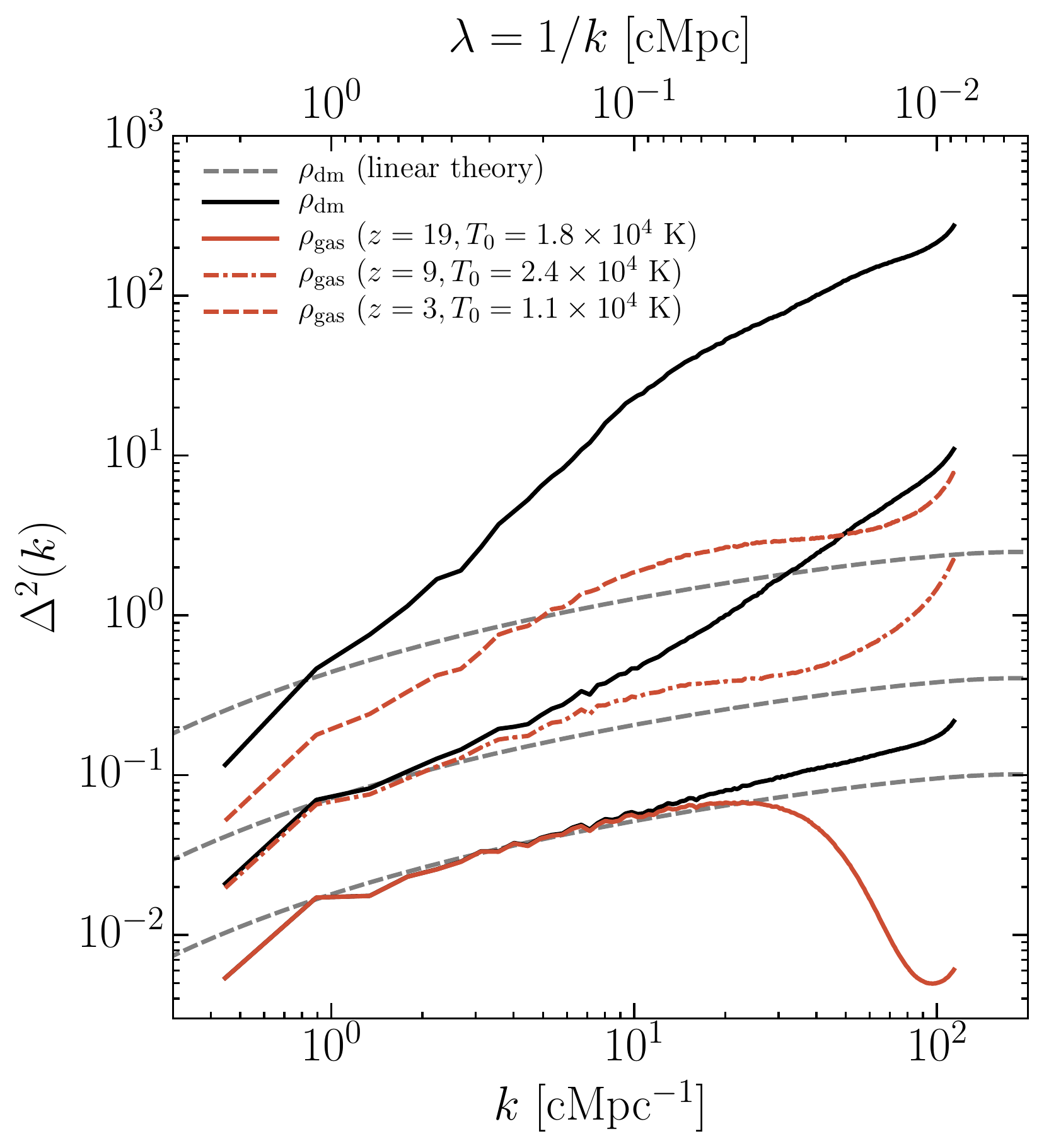}
    % ps.py (case 10)
  \end{center}
  \caption{Evolution of the gas density power spectrum (red curves)
    from redshift 19 to 3 in a simulation in which the photoheating
    rate is enhanced so that the gas temperature is $T_0\sim 10^4$~K
    at $z=19$.  Also shown are the nonlinear (solid black curves) and
    linear (dashed grey curves) dark matter density power spectra at
    each redshift.  The gas power spectrum shows a Gaussian cut-off as
    expected from Equation~(\ref{eqn:fullsoln}) at $z=19$, but this
    cut-off vanishes at low redshifts.}
  \label{fig:ps_hothighz_evolution}
\end{figure}

For all simulations, we take snapshots of particle positions,
velocity, temperature, and other properties.  To calculate power
spectra, we grid the relevant particles to create a density field,
using a cloud-in-cell (CIC) scheme, taking into account the smoothing
lengths of SPH particles.  After calculating the power spectrum, we
deconvolve the CIC kernel, ignoring small errors due to aliasing on
the smallest scales \citep{2008ApJ...687..738C}.

\begin{figure}
  \begin{center}
    \includegraphics*[width=\columnwidth]{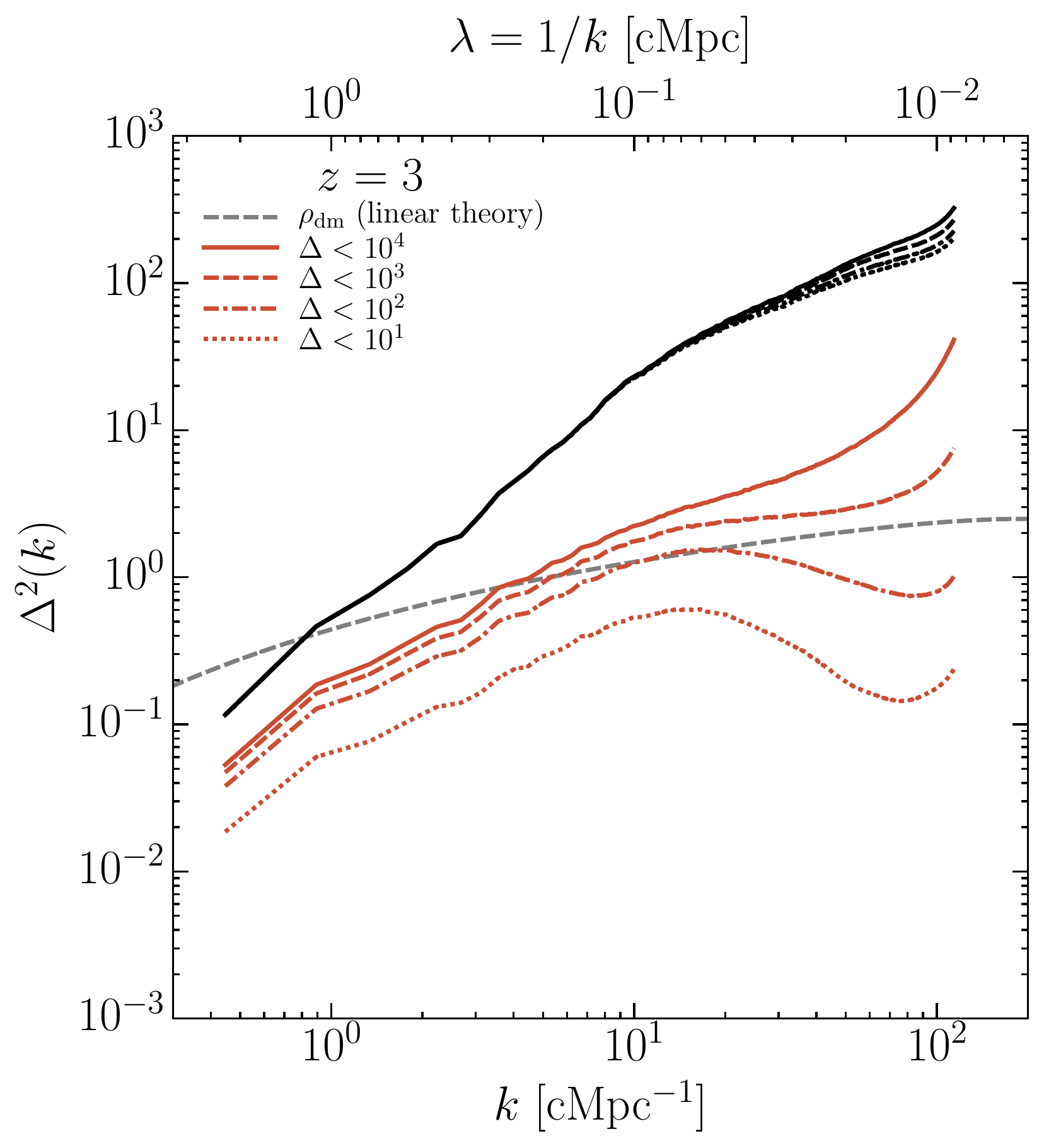}
    % ps.py case 21
  \end{center}
  \caption{Gas density power spectrum with different star formation
    thresholds.  The solid, dashed, dot-dashed, and dotted red curves
    show the gas density power spectra when the star formation
    threshold is set to $10^4$, $10^3$, $10^2$ and $10$, respectively.
    The black curves show corresponding dark matter power spectra.
    Increasing the star formation threshold removes high density gas,
    which reduces small scale power.}
  \label{fig:psthresh_onlydensity}
\end{figure}

Ly$\alpha$ forest spectra are created from the particle distributions
of each snapshot, accounting for the redshift-space distortions caused
by peculiar velocities and thermal broadening.  These are computed by
accounting for the contributions to density, temperature and velocity
of SPH particles closer to the sightline than their smoothing length,
Doppler shifts due to the bulk velocity, and the thermal broadening of
the Ly$\alpha$ absorption line.  This procedure results in the
Ly$\alpha$ flux as a function of wavelength or equivalently time or
distance.  Following the standard approach, we then rescale the UV
background intensity so that the mean flux of these extracted spectra
matches the observed mean flux at the respective redshift
\citep{2010MNRAS.404.1281P}.  We use the Ly$\alpha$ flux measurements
reported by \citet{2008ApJ...688...85F} for this purpose.

\section{Characterizing the Pressure Smoothing Scale}
\label{sec:results}

We now turn to the results of our simulations.  Our first goal is to
compare the simulations with the linear theory predictions of
Equations~(\ref{eqn:bidavid}) and (\ref{eqn:fullsoln}).

\subsection{Absence of a Cut-Off in the Gas Density Power Spectrum}

Figure \ref{fig:ps} shows the dimensionless power spectra of gas and
dark matter density contrasts ($\delta=\rho/\bar\rho-1$; for gas we
use $\bar\rho=\Omega_b\rho_\mathrm{cr}$) at $z=3$ from our fiducial
simulation.  The dark matter density power spectrum increases steeply
towards small scales.  The amplitude of the gas density power spectrum
is suppressed relative to the dark matter at all scales probed by the
simulation.  Note that the enhancement in the dark matter power
relative to the linear theory prediction indicates that scales of
several Mpc are already nonlinear at this redshift.  Therefore, it is
clear that the pressure smoothing scale, which is expected to be of order 100
kpc, is already evolving non-linearly.

However, it is striking that in Figure \ref{fig:ps} the gas density
power does not exhibit a cut-off on any scale.  This is very different
from the linear theory expectations discussed in
Section~\ref{sec:past}, which suggested a quadratic or Gaussian
cut-off in the gas power spectrum at the pressure smoothing scale according to
Equations~(\ref{eqn:bidavid}) and (\ref{eqn:fullsoln}).  This
discrepancy is highlighted in Figure \ref{fig:ps}, where the
predictions of Equations~(\ref{eqn:bidavid}) and (\ref{eqn:fullsoln})
are also shown for an arbitrarily chosen pressure smoothing scale.  The absence
of a cut-off in the gas power spectrum shows that the linear theory
formalism of Section~\ref{sec:past} does not describe the physics of
pressure smoothing on non-linear scales at $z=3$.  We observe similar
behavior, i.e., an absence of a cut-off in the gas density power
spectrum, in our simulations over the complete redshift range probed
by the Ly$\alpha$ forest ($z\sim 2$--$5$).

To understand this large difference between the gas density power
spectrum of Figure~\ref{fig:ps} and the linear theory predictions of
Equations~(\ref{eqn:bidavid}) and (\ref{eqn:fullsoln}), it is useful
to consider the gas and dark matter density power spectra at redshifts
where density perturbations are still linear.  We therefore look at
the power spectra at $z= 19$ in Figure~\ref{fig:ps_z19_multipleT}.  At
this high redshift, density perturbations are linear at most scales
resolved in our box as indicated by the agreement of the dark matter
density power spectrum with that predicted by linear theory in
Figure~\ref{fig:ps_z19_multipleT}. Scales $\sim 100$ kpc comparable to
the pressure smoothing scale are quasi-linear, and the assumptions behind
Equation~(\ref{eqn:fullsoln}) are thus still valid. We would therefore
expect the gas power spectrum to exhibit a Gaussian cut-off.  However,
we face a different problem here at $z=19$, namely that the gas
temperature in the fiducial simulation is $T_0=6$~K as reionization
does not occur until $z=6$.  At this low temperature, the filtering
scale (Equation~(\ref{eqn:kf})) is expected to be $\sim 2$ kpc from
Equation~(\ref{eqn:jeans}), much smaller than the smallest scale
resolved in the simulation.  As a result, there is no clear evidence
for a cut-off in the gas density power spectrum.  Indeed, in
Figure~\ref{fig:ps_z19_multipleT}, the gas and dark matter power
spectra from the fiducial simulation agree at almost all scales,
verifying our expectation that at low temperatures the tiny filtering
scale $\sim 2$ kpc implies that gas follows dark matter for all scales
resolved by the simulation.

To remedy this, we consider three other simulations for which we
artificially enhance the H~\textsc{i} photoheating rate between $z\sim
6$ and $30$.  Specifically, at redshifts $z<6$ the photoheating rate
is equal to that in the fiducial simulation, but it is set to be much
higher ($10^{-22}$--$10^{-24}$ erg$/$s) at $z > 6$.  These simulations
with enhanced heating at high redshift can be thought of as ``early
reionization'' models.  Due to this enhanced heating the gas
temperature at $z=19$ is $T_0\sim 10^4$~K in these simulations, which
increases the pressure smoothing scale of Equation~(\ref{eqn:kf}) to be well
above the spatial resolution of our simulation so that any cut-off
will now be visible.  Figure \ref{fig:ps_z19_multipleT} shows the gas
density power spectra at $z=19$ from these three simulations with
various enhanced heating rates and gas temperatures.  The gas power is
seen to drop precipitously between scales of 50 and 100 kpc, and the
dotted curve shows that this cut-off has a Gaussian form, in
accordance with Equation~(\ref{eqn:fullsoln}).  Moreover, as the gas
temperature $T_0$ increases, the cut-off is seen to move towards
larger scales as expected for the pressure smoothing scale from
Equation~(\ref{eqn:kf}).  Thus the classical pressure smoothing scale result
appears to be valid at very high redshifts, when the $\sim 100$ kpc
pressure smoothing scale is quasi-linear, but seems to break down at $z\sim 3$
when scales $\sim 100$ kpc become highly nonlinear.

\begin{figure}
  \begin{center}
    \includegraphics*[width=\columnwidth]{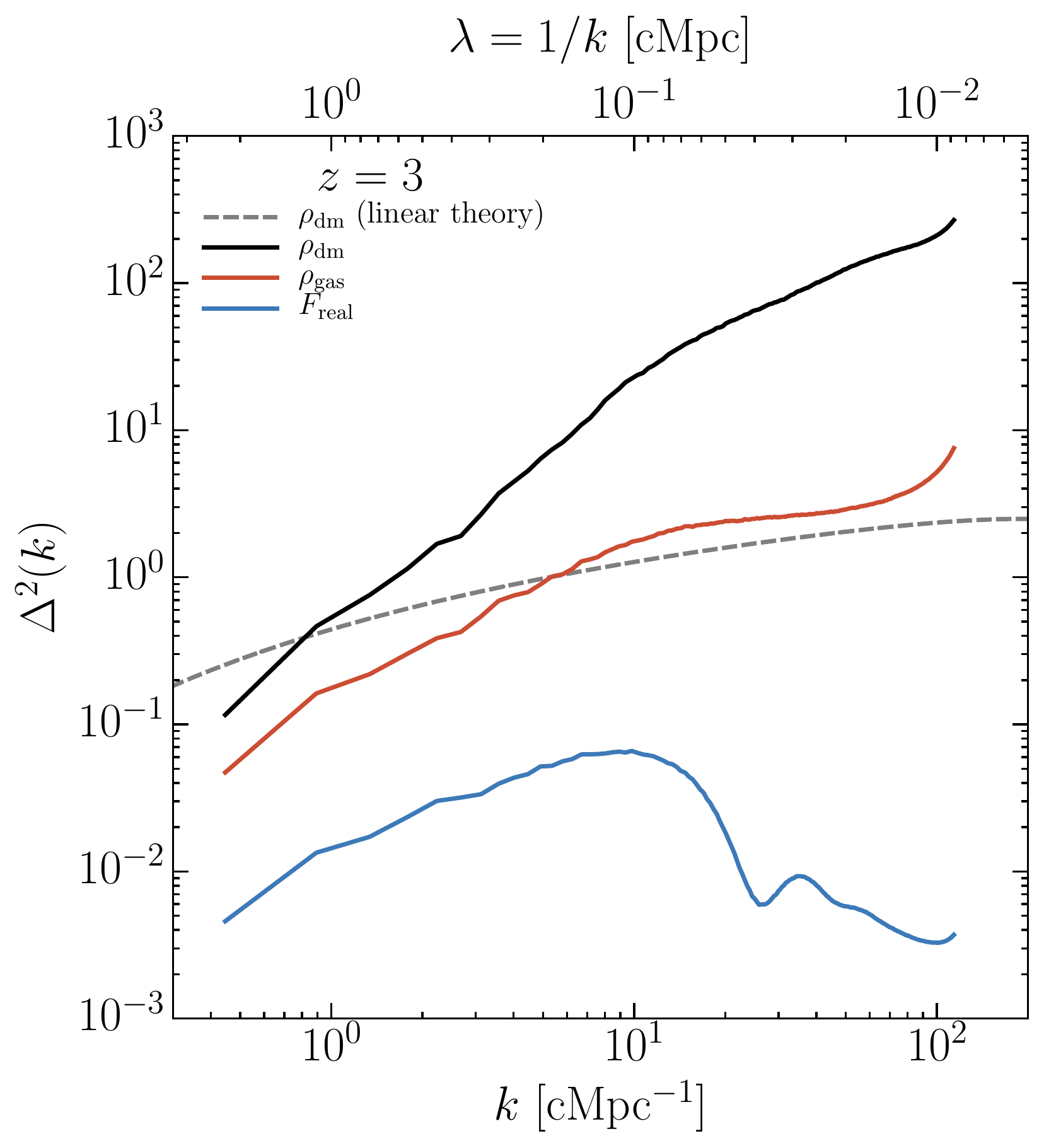}
    % ps.py case 14 
  \end{center}
  \caption{Power spectra of the dark matter density (black), gas
    density (red), and $F_\mathrm{real}$ (blue) fields at $z=3$ from
    our fiducial simulation.  The dashed grey curve shows the matter
    power spectrum prediction for linear growth of perturbations.}
  \label{fig:freal}
\end{figure}

\begin{figure}
  \begin{center}
    \includegraphics*[width=\columnwidth]{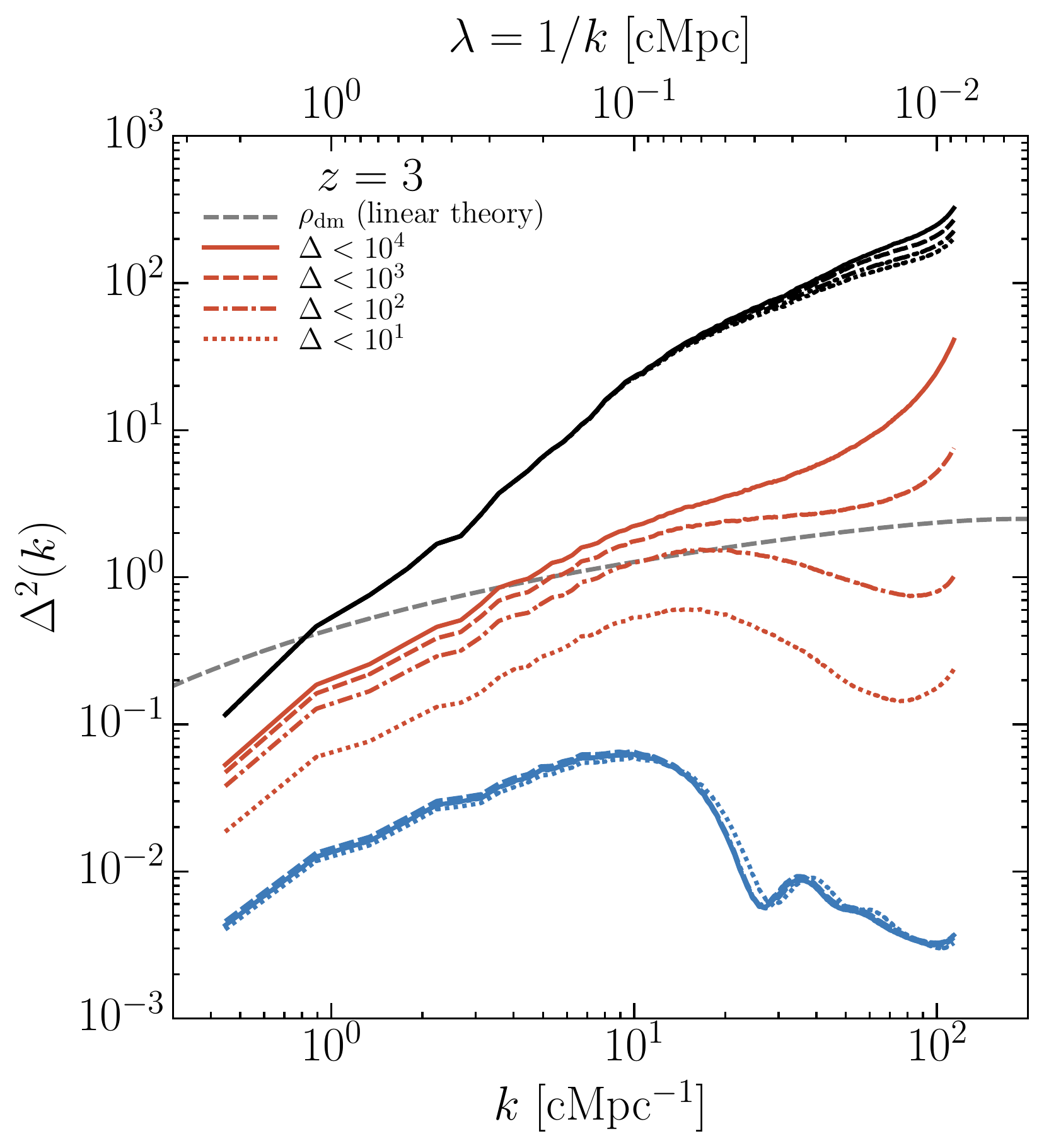}
    % ps.py case 0
  \end{center}
  \caption{$F_\mathrm{real}$ power spectra with different star
    formation density thresholds.  The threshold, which acts as a
    crude model of galaxy formation, affects the gas density power
    spectra dramatically but $F_\mathrm{real}$ is insensitive to it.}
  \label{fig:psthresh}
\end{figure}

\begin{figure*}
  \begin{center}
  \begin{tabular}{cc}
    \includegraphics*[width=\columnwidth]{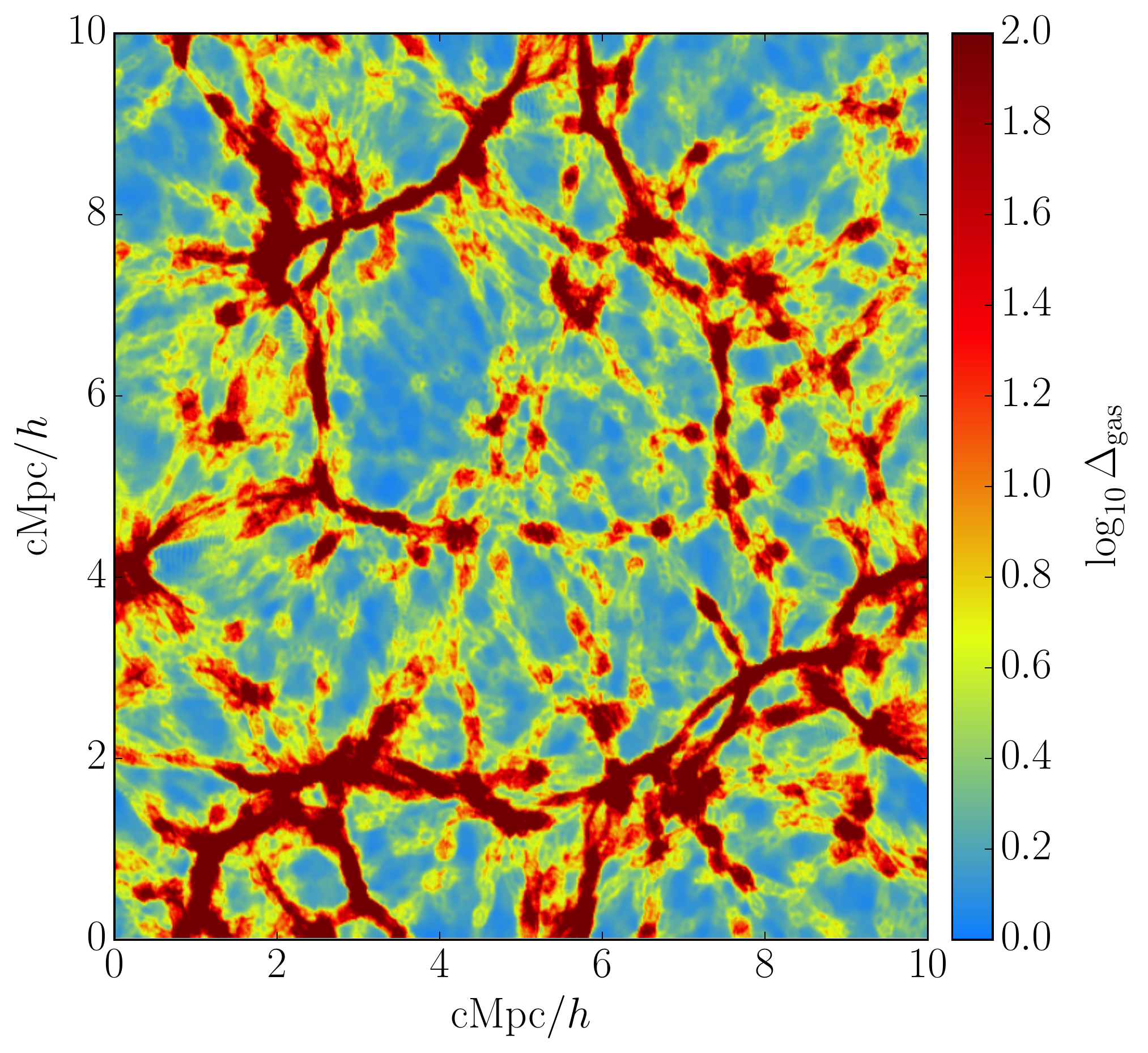} &
    % slice_all.py 
    \includegraphics*[width=\columnwidth]{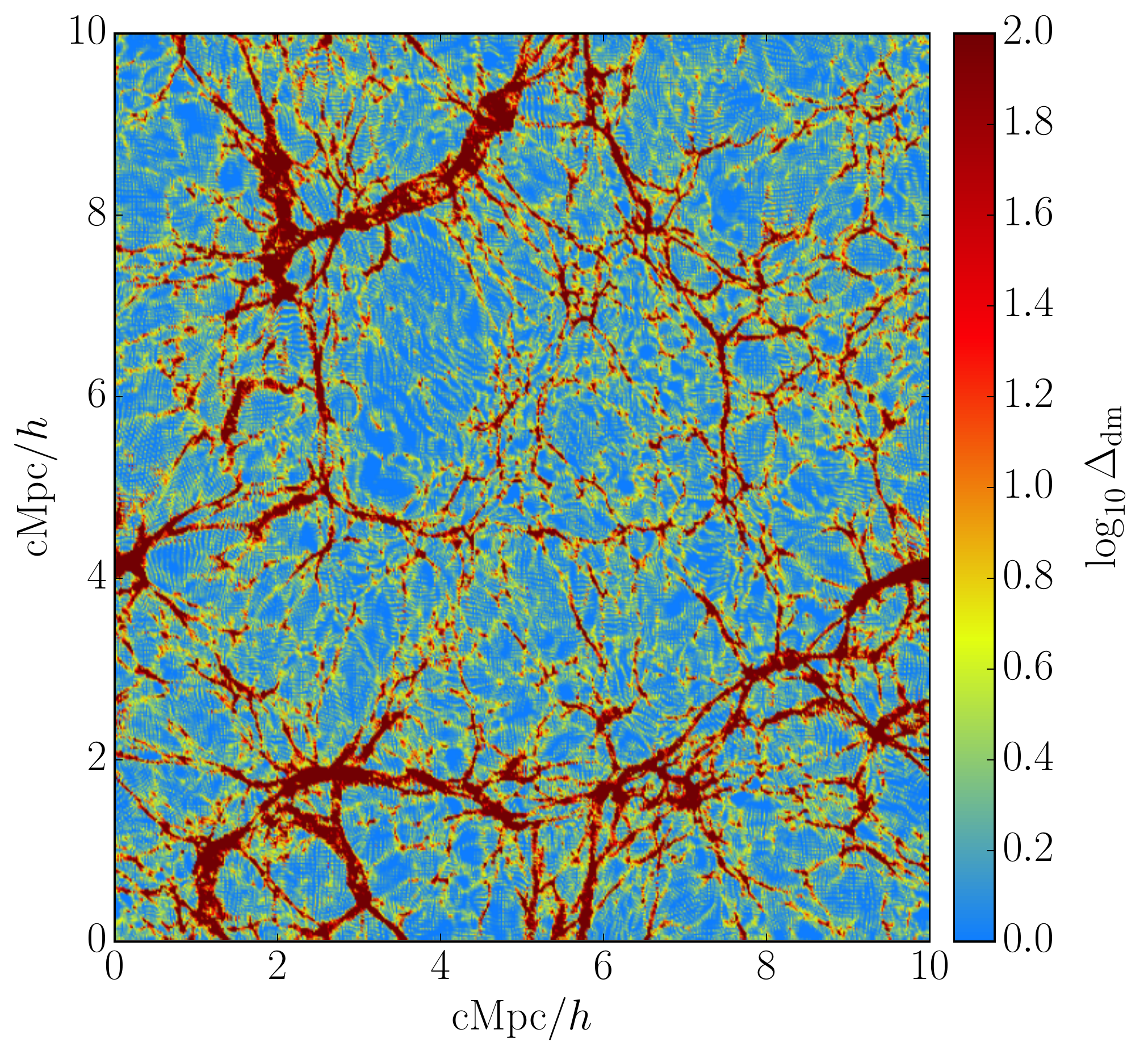} \\
    % slice_all.py 
    \includegraphics*[width=\columnwidth]{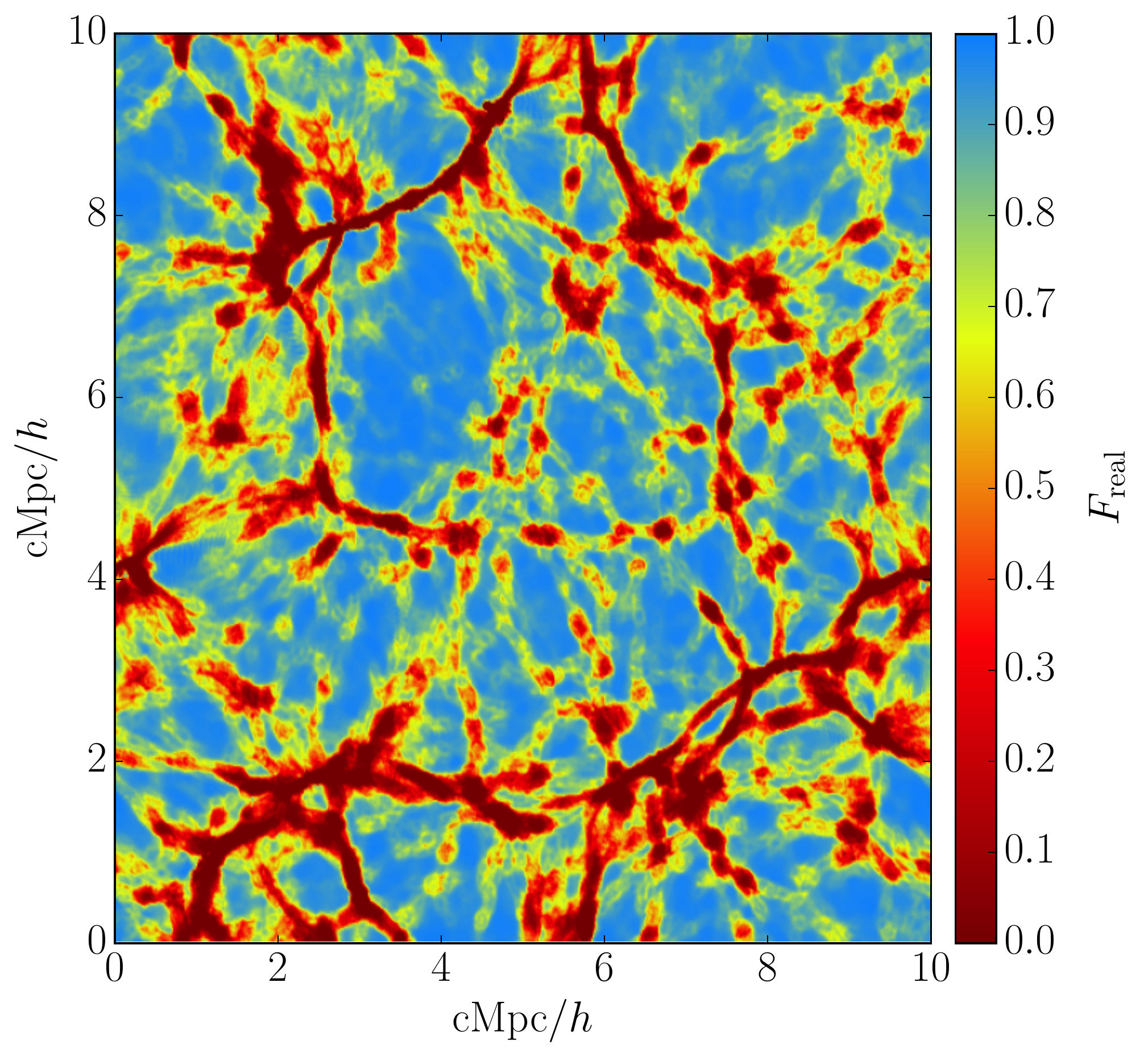} & 
    % slice_all.py 
    \includegraphics*[width=\columnwidth]{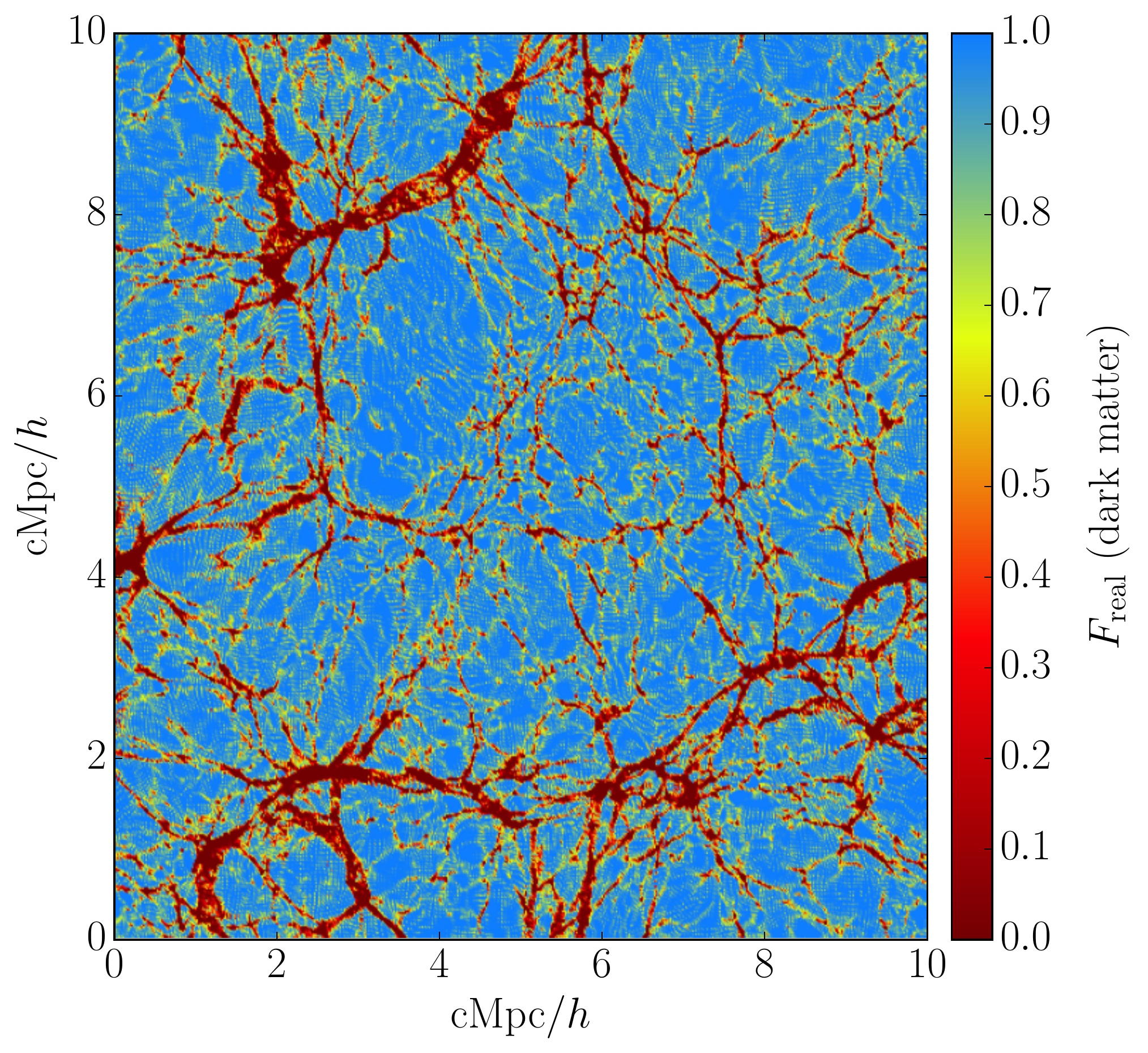} \\ 
    % slice_all.py 
  \end{tabular}
  \end{center}
  \caption{Slices of gas (top left) and dark matter (top right)
    density, and of the $F_\mathrm{real}$ distribution calculated from
    gas (bottom left) and dark matter (bottom right) distributions at
    $z=3$ in our fiducial simulation.  Density at each point is an
    average on a cubical cell approximately 20 ckpc$/h$ on a side.
    Each slice has a thickness of one cell length.  Note that the
    color scale in panels showing $F_\mathrm{real}$ is inverted.  The
    mapping from density to $F_\mathrm{real}$ reduces high density
    regions to zero.}
  \label{fig:all_slices}
\end{figure*}

This breakdown of the linear theory pressure smoothing scale picture is clearly
indicated in Figure~\ref{fig:ps_hothighz_evolution}, which shows the
evolution from $z=19$ to $3$ of gas and dark matter density power
spectra in one of the simulations with enhanced photoheating at high
redshifts.  Comparison with the linear theory matter power spectrum
(gray dashed curves) indicate that scales $\sim 100$ kpc are already
highly nonlinear by $z=9$.  Hence one sees that at $z=19$, when the
pressure smoothing scale is quasi-linear, the gas power spectrum shows a clear
Gaussian cut-off at $50$ comoving kpc, whereas at lower redshifts when
this scale is highly nonlinear this cut-off vanishes.  In what
follows, we will explain why the cut-off disappears at low redshifts
and will present a method to reveal it.

\subsection{The Role of High Density Gas}
\label{sec:high_density_regions}

As we saw above in Figure~\ref{fig:ps_z19_multipleT}, the gas density
power spectrum does not exhibit any cut-off in power at scales $\sim
100$ kpc at relatively low redshifts ($z\sim 3$) in our fiducial
simulation.  At these redshifts and small scales, the power is
completely dominated by highly overdense and non-linear collapsed
structures, namely halos. Although these structures occupy a tiny
volume fraction, they nevertheless have a large impact on the power
spectrum.  Recall that galaxy formation is simplified in our fiducial
simulation: all gas particles with overdensity greater than $10^3$ and
temperature less than $10^5$~K are converted into stars.  This results
in a stellar density parameter $\Omega_*=0.0075$ at $z=3$, which is
already factor of three higher than the measured value of
$\Omega_*=0.0027$ at $z=0$ \citep{2004ApJ...616..643F}.  We now rerun
this simulation, with the same initial conditions, but with modified
star-formation overdensity thresholds of $10^4$, $100$, and $10$. At
$z=3$ we find that thresholds of $10^4$, $10^3$, $100$, and $10$,
result in $\Omega_*=0.0066$, $0.0075$, $0.0092$, and $0.0145$,
respectively.  As we reduce the star formation density threshold, we
form more stars but at the same time we remove more high density gas
from our simulation and the power spectrum analysis.

\begin{figure*}
  \begin{center}
    \includegraphics*[width=\textwidth]{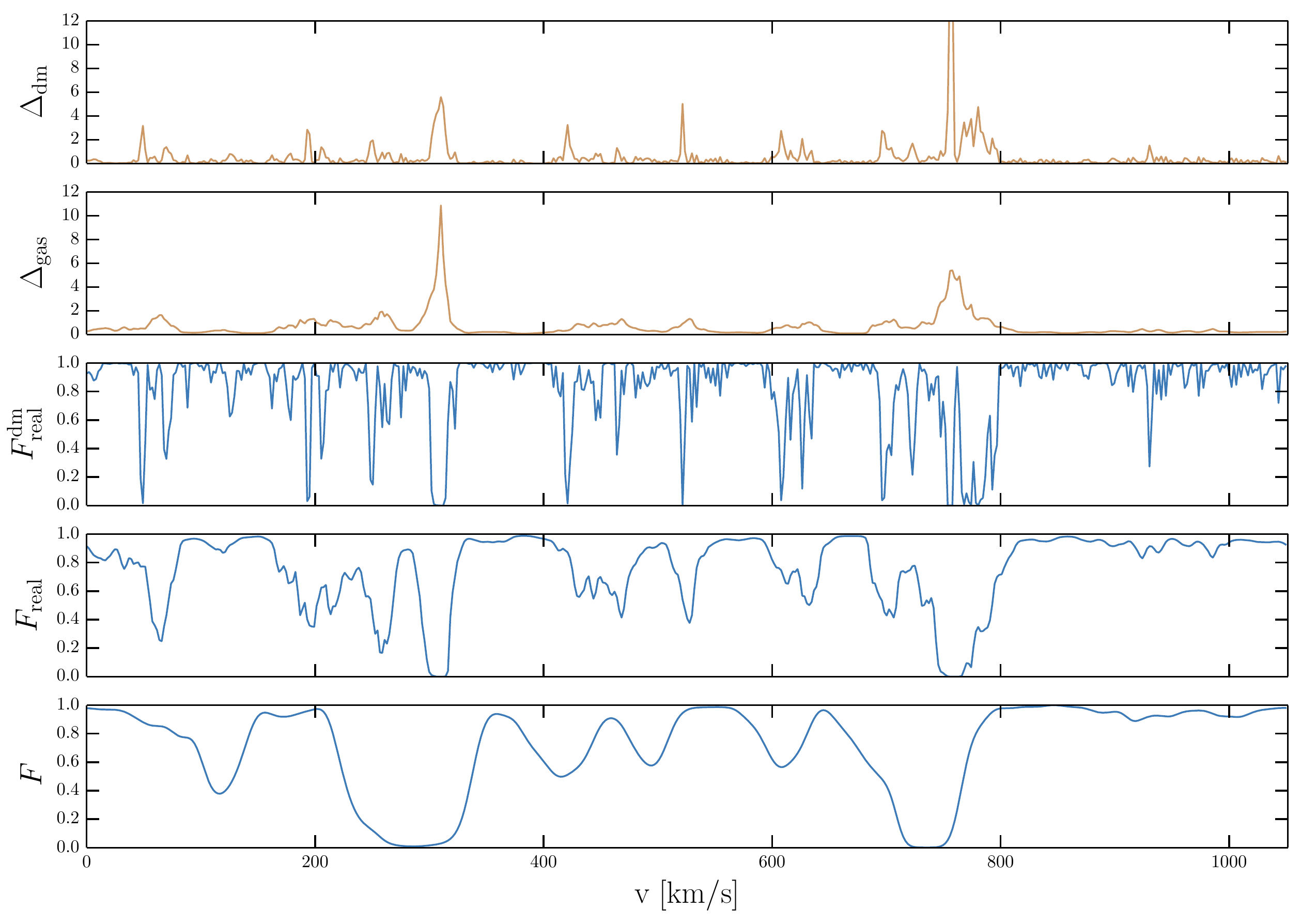}
    % ps.py case 17
  \end{center}
  \caption{Distribution of various quantities along a randomly-chosen
    line of sight through our fiducial simulation.  From top to
    bottom, panels show the dark matter overdensity, gas overdensity,
    the distribution of $F_\mathrm{real}$ calculated under the
    assumption that gas follows dark matter at all scales (see
    Section~\ref{sec:viz}), and the distribution of $F_\mathrm{real}$
    calculated from the gas distribution in the simulation.  The
    bottom panel shows the Ly$\alpha$ flux along the line of sight,
    where the effect of longitudinal thermal line-broadening and
    redshift-space distortions are clearly seen.}
  \label{fig:skewers}
\end{figure*}

Figure \ref{fig:psthresh_onlydensity} shows the resulting gas density
power spectra for these different star-formation density thresholds.
As the density threshold is progressively decreased, and more high
density gas is removed from the simulation, we see that the power on
scales smaller than $\lesssim 100\,$kpc is increasingly attenuated,
and the overall amplitude of the power on larger scales is also
reduced.  This behavior is not unexpected, and can be understood in a
halo model picture \citep{2002PhR...372....1C}.  As dense gas above
the star-formation threshold is removed, we effectively mask dense gas
in collapsed haloes in the simulation volume. The truncation of the
small scale power arises from the removal of this gas, which can be
thought of as suppressing the `one-halo' term of the gas power
spectrum. However, part of the large scale power of the gas density
also arises from the clustering of these halos, i.e. the `two halo
term', hence masking haloes results in a reduction of large scale
power as well.  Therefore on large scales, the net effect of the
star-formation threshold is to introduce a linear bias, such that the
power spectra for different star-formation thresholds in
Figure~\ref{fig:psthresh_onlydensity} are all parallel to each other
at small $k$ values.  Figure~\ref{fig:psthresh_onlydensity} thus shows
us that the gas density power spectrum is indeed strongly influenced
by nonlinear structure.  More importantly, as the highest density
regions are removed, the gas distribution begins to show a hint of a
cut-off in power.  This is seen most clearly in the $\Delta<10$ curve
in Figure~\ref{fig:psthresh_onlydensity}.

In other words, to recover the signature of pressure smoothing in the
moderately overdense IGM ($\Delta\lesssim 10$), we need to remove the
influence of gas in high-density regions, which would otherwise
dominate the small-scale power.  One way of doing this is by applying
an ad hoc cut-off in gas density as done in
Figure~\ref{fig:psthresh_onlydensity}.  However, the choice of this
density threshold is arbitrary---it would need to be made
redshift-dependent, as the amount of high-density gas that has
collapsed to form stars increases at later times, and is thus
dependent on the galaxy formation model in the simulation, which we
have crudely parameterized here with a simple star-formation density
threshold. However, even in more sophisticated simulations which
attempt to model detailed baryonic process in galaxies such as
star-formation and feedback using a combination of higher resolution
and sub-grid modeling \citep[e.g.,][]{2010MNRAS.402.1536S,
  2014Natur.509..177V, 2015arXiv150101311C}, our findings indicate
that the the small scale baryon power spectrum will be influenced by
the details of galaxy formation and the specific sub-grid
implementation.  Therefore the challenge is to devise a more robust
method of removing the influence of high density gas on small scale
power.

\subsection{Real-Space Flux Reveals Pressure Smoothing in Baryons} 

In this section we introduce a better method for revealing the
three-dimensional cut-off in the baryon power of the moderate
overdensity IGM.  Given the three-dimensional, statistically isotropic
gas density field, we calculate the \emph{real-space} Ly$\alpha$ flux
at each point in the box.  This quantity, which we denote by
$F_\mathrm{real}$, is defined by
\begin{equation}
  F_\mathrm{real} \equiv \exp(-\tau_\mathrm{real}),
  \label{eqn:freal}
\end{equation}
where the real-space Ly$\alpha$ absorption optical depth is given by 
\begin{equation}
  \tau_\mathrm{real} = \frac{3\lambda_\alpha^3\Lambda_\alpha}{8\pi
    H(z)}n_\mathrm{HI}.
  \label{eqn:tau}
\end{equation}
Here $\lambda_\alpha=1216$ \AA\ is the rest-frame Ly$\alpha$
wavelength, $\Lambda_\alpha$ is the Einstein A coefficient (also
written as $A_{10}$ and can be written in terms of an oscillator
strength), $H(z)$ is the Hubble constant, and $n_\mathrm{HI}$ is the
number density of neutral hydrogen. The definition in
Equation~(\ref{eqn:tau}) is identical to the Gunn-Peterson formula
used to compute the Ly$\alpha$ forest optical depth, except that the
convolution integral which accounts for the redshift-space effects of
the peculiar velocity field and thermal line broadening has not been
included.  Indeed, we have chosen to define a statistic in real space
precisely because these redshift-space effects, specifically the
line-of-sight smoothing due to thermal broadening, are in fact
degenerate with the three-dimensional real-space Jeans smoothing that
we aim to study. Hence, $F_\mathrm{real}$ is a three-dimensional,
statistically isotropic field.

The neutral hydrogen density is given by
\begin{equation}
  n_\mathrm{HI} =
  \frac{\alpha_R(T)n^2_\mathrm{H}}{\Gamma_\mathrm{HI}},
\label{eqn:nhi}
\end{equation}
where $n_\mathrm{H}=0.76\rho_\mathrm{gas}$ is the total hydrogen
density, $\Gamma_\mathrm{HI}$ is the H~\textsc{i} photoionization
rate, $\alpha_R(T)\propto T^{-0.7}$ is the hydrogen recombination
rate.  Therefore, in a highly ionized IGM where the gas lies on a
well-defined temperature density relation $T\propto \rho^{\gamma-1}$
(see Figure~\ref{fig:trho}), the neutral hydrogen number density and
the optical depth in Equation~(\ref{eqn:tau}) are proportional to a
power of the total gas density $\rho^{2-0.7(\gamma-1)}$.  (When all
gas is assumed to lie on the temperature-density relation,
Equation~(\ref{eqn:nhi}) constitutes the Fluctuating Gunn-Peterson
Approximation (FGPA).)  Thus, $F_\mathrm{real}$ is proportional to the
\emph{exponential} of this power of the gas density, and it
exponentially suppresses the contribution of the high density regions.
This is of course another way to say that the high density regions
have large optical depth and therefore result in completely saturated
Ly$\alpha$ absorption, i.e., have $F_\mathrm{real}=0$.

$F_\mathrm{real}$ is not a directly observable field, but it can
nevertheless be constrained by close quasar pair observations
\citep{2013ApJ...775...81R}.  The phase angle differences between
homologous line-of-sight Fourier modes in quasar pair spectra can be
written as
\begin{equation}
  \theta_{12}(k)=\cos^{-1}\left[\frac{\mathrm{Re}\left(F_{1k}^*F_{2k}\right)}{|F_{1k}F_{2k}|}\right],
\end{equation}
where $F_1(k)$ and $F_2(k)$ are Fourier amplitudes of the two quasars
in a pair.  \citet{2013ApJ...775...81R} showed that the probability
distribution of these phase differences $\theta_{12}(k)$ is sensitive
to a smoothing cut-off in the underlying field.  For modes with
wavelength larger than the pair separation $r_{\perp}$, $k\ll
k_{\perp}\sim 1\slash r_{\perp}$, the mean value of $\theta_{12}(k)$
can be written as \citep{2013ApJ...775...81R}
\begin{equation}
  \langle \cos\theta(k,r_{\perp})\rangle \approx
  \frac{\int_k^{k_{\perp}}dk^\prime k^\prime
    J_0(r_\perp\sqrt{{k^\prime}^2 -
      k^2})P_{F_\mathrm{real}}(k^\prime)}{\int_k^\infty dk^\prime
    k^\prime P_{F_\mathrm{real}}(k^\prime)}
\end{equation}
where $J_0$ is the Bessel function of zeroth order and
$P_{F_\mathrm{real}}$ is the three-dimensional power spectrum of
$F_\mathrm{real}$.  Thus by measuring the phase angle probability
distribution function in close quasar pairs the $F_\mathrm{real}$ can
be directly constrained.

Note that in the construction of the $F_\mathrm{real}$ field using
Equation~\ref{eqn:freal}, we use the rescaled photoionization rate
$\Gamma_\mathrm{HI}$ that produces the correct observed mean
\emph{redshift-space} flux as discussed in Section~\ref{sec:sims}
above.  In this approach, we rescale the UV background intensity so
that the mean redshift-space flux of these extracted spectra matches
the observed mean redshift-space flux at the respective redshift
\citep{2008ApJ...688...85F}.  When this rescaled background is used in
calculating the real-space flux $F_\mathrm{real}$, its mean value is
not exactly the same as the mean redshift-space flux, although the
differences are very small. For instance, in our fiducial simulation,
the mean redshift-space flux at $z=3$ is 0.680 after rescaling the UV
background, while the corresponding mean value of $F_\mathrm{real}$,
using this same background, is 0.679.  Equations~(\ref{eqn:tau}) and
(\ref{eqn:nhi}) indicate that rescaling the UV background to match the
observed mean flux values is tantamount to fixing the range of gas
densities probed by the flux field.  As such, the very close agreement
between the mean of the redshift space $F$ and real-space
$F_\mathrm{real}$ flux indicates that $F_\mathrm{real}$ and $F$ probe
nearly identical gas densities.

In Figure \ref{fig:freal} we now look at the three-dimensional power
spectrum of the distribution of $F_\mathrm{real}$ at $z=3$ in our
fiducial simulation (recall that unlike the observed Ly$\alpha$ flux,
$F_\mathrm{real}$ is statistically isotropic).  Figure \ref{fig:freal}
also shows the gas and dark matter density power spectra for
comparison.  The most conspicuous feature of the $F_\mathrm{real}$
power spectrum in Figure~\ref{fig:freal} is that it exhibits a
small-scale cut off, located at around $k=20~h/$Mpc (about $\lambda =
35$ comoving kpc)\footnote{Note that we choose to define the
  upper-axis as $1/k$ instead of $2\pi/k$ because for a Gaussian
  smoothing the Fourier space cut off is inverse of that in the real
  space.  As we see below, the cut off in $F_\mathrm{real}$ has a
  Gaussian form.}.  This is reminiscent of the Gaussian filtering
scale cut-off that we expect from linear theory.  There is also an
overall reduction in the amplitude of the $F_\mathrm{real}$ power
relative to that of the gas density.  This reduced power level is
similar to that seen when we applied a star formation density
threshold in Figure~\ref{fig:psthresh_onlydensity}, and, which we
argued can also be understood in the halo model picture.  Generically,
a nonlinear transformation of the density field, such as
$F_\mathrm{real}$, will have a power spectrum that has a different
shape than that of the density, but for small $k$ where the Fourier
amplitudes are small, the transformation can be linearized such that
the power will appear to be a linear bias rescaling of that of the
original density field.

Thus $F_\mathrm{real}$ provides a natural way of removing the
influence of high density regions in our analysis, and reveals the
small-scale structure of the IGM---in the form of a sharp cut-off in
the power at a certain scale---that is presumably due to pressure
smoothing.  Furthermore, this definition of $F_\mathrm{real}$ is
robust against arbitrary choices of galaxy formation prescriptions in
our simulations. Recall that, as discussed in
Section~\ref{sec:high_density_regions}, we can think of variation in
the star formation threshold as variation in our (crude) galaxy
formation prescriptions.  The red curves in Figure~\ref{fig:psthresh}
illustrate that the gas density power spectrum varies dramatically as
the star formation density thresholds is varied from $10^4$ to $10$.
As the threshold is decreased, power at small scales is considerably
reduced, but a clear signature of the pressure smoothing is not
apparent from these curves.  The removal of clustered high-density
regions, also reduces the large-scale power as a linear-bias
rescaling.  Although the density power varies dramatically depending
on the details of how galaxy formation is treated, the blue curves in
Figure~\ref{fig:psthresh} show that the shape of the $F_\mathrm{real}$
power spectrum is invariant to these changes.  Thus the
$F_\mathrm{real}$ field as defined in Equation~(\ref{eqn:freal})
provides an unambiguous way of characterizing the cut-off, valid at
all relevant redshifts regardless of the galaxy formation details used
in the simulation.

\subsection{Visualizing the $F_\mathrm{real}$ Field}
\label{sec:viz}

Figure~\ref{fig:all_slices} shows a slice of the $F_\mathrm{real}$
field at $z=3$ from our fiducial simulation in comparison with a slice
of the gas density field.  There is an almost one-to-one
correspondence between gas density and $F_\mathrm{real}$.  This
correspondence results from ignoring redshift-space distortions due to
peculiar velocities and thermal broadening.  The value of
$F_\mathrm{real}$ is highest ($\sim 1$) in regions with lowest gas
density, whereas all high density filaments and haloes are mapped to
$F_\mathrm{real}=0$.  This is exactly the kind of suppression of high
densities required to reveal the structure of the low-density IGM.
Note that unlike in Figure~\ref{fig:gas_and_dm_slices}, we have chosen
a linear color scale in the gas density plot to highlight the
saturation of $F_\mathrm{real}$ values at high densities. 

For the sake of illustration, we also apply the transformation encoded
in $F_\mathrm{real}$ on the dark matter density field.  This should be
understood as the $F_\mathrm{real}$ field in the limit where the
baryons directly trace the dark matter density, but nevertheless
adhere to the nonlinear transformation in Equation~(\ref{eqn:nhi}).
This is achieved by ``gassifying'' dark matter, i.e., mapping the dark
matter density field to a pseudo-gas density field through
\begin{equation}
  \rho_\mathrm{gas} =
  \rho_\mathrm{dm}\left(\frac{\Omega_b}{\Omega_m-\Omega_b}\right), 
\end{equation}
which traces the dark matter at all scales.  In other words, as the
dark matter is cold and pressureless and exhibits significant
small-scale power, this pseudo-gas distribution has no pressure
smoothing scale.

The neutral hydrogen number density, $n_\mathrm{HI}$, is then
calculated by imposing the same temperature-density relation on these
pseudo-gas particles as our fiducial simulation, i.e., by using FGPA.
A slice of the resulting $F^\mathrm{dm}_\mathrm{real}$ field
calculated in this fashion is shown in the lower right panel of Figure
\ref{fig:all_slices}.  Its structure is markedly different from that
of the $F_\mathrm{real}$ field calculated from the gas distribution.
The $F^\mathrm{dm}_\mathrm{real}$ has a morphology very similar to the
dark matter density distribution, and because it lacks a pressure
smoothing scale cut-off, has a significant amount of small-scale
structure. In contrast the pressure smoothing of the gas distribution
makes the $F_\mathrm{real}$ field smoother and more diffuse.  Although
somewhat contrived, $F^\mathrm{dm}_\mathrm{real}$ is helpful in better
understanding the $F_\mathrm{real}$ transformation of the density
field.

These points are further illustrated in Figure~\ref{fig:skewers},
which compares the $F_\mathrm{real}$ field along an example line of
sight through our fiducial simulation to various other quantities at
$z=3$.  The upper two panels of this figure show the dark matter and
gas density field along the line of sight.  As in
Figure~\ref{fig:all_slices}, we see that the gas density field is a
smoothed version of the dark matter density field.  The third and
fourth panels of Figure~\ref{fig:skewers} show the
$F^\mathrm{dm}_\mathrm{real}$ and $F_\mathrm{real}$ fields, calculated
from the dark matter and gas density fields respectively.  The mapping
to real space flux makes the smoothness of the gas density field
relative to the dark matter density field much more apparent.  The
last panel shows the observed, redshift-space Ly$\alpha$ flux, $F$,
which includes the effects of peculiar velocities and thermal
broadening.  As a result, the absorption features in $F$ are shifted
because of peculiar velocities, and their widths now reflect the
impact of Hubble flow and are much smoother because of thermal line
broadening. The comparison between $F_\mathrm{real}$ and $F$ is
striking, and illustrates that the intrinsic small-scale structure of
the gas in real space, which is determined by pressure smoothing, is
completely hidden by redshift space effects.

\begin{figure}
  \begin{center}
    \includegraphics*[width=\columnwidth]{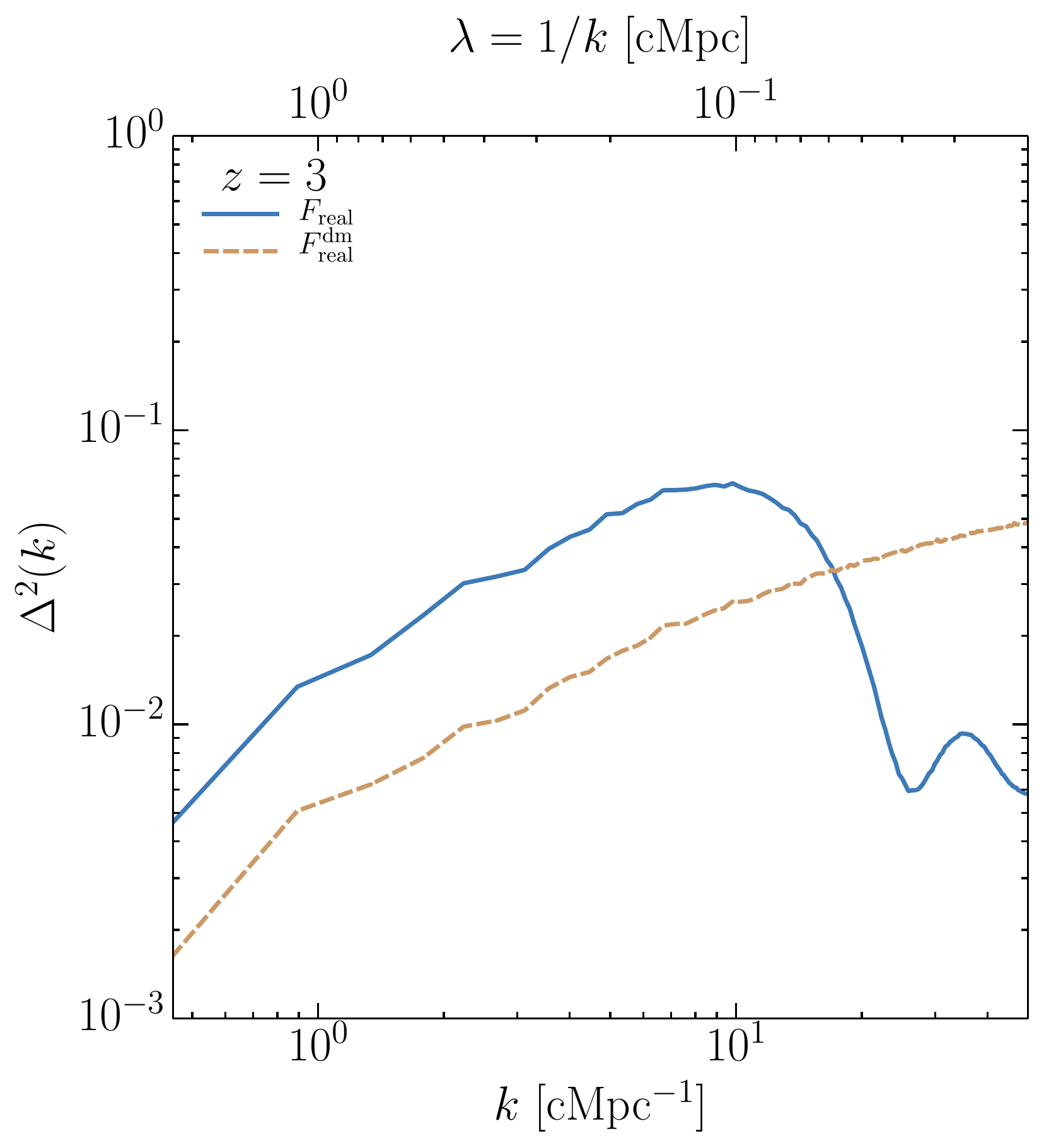}
    % ps.py case 17
  \end{center}
  \caption{Power spectrum of the $F_\mathrm{real}$ distribution when
    calculated under the assumption that gas follows dark matter at
    all scales (dashed curve), compared to the power spectrum of
    $F_\mathrm{real}$ calculated from the actual gas distribution in
    the simulation (solid curve).}
  \label{fig:freal_fromgasdm}
\end{figure}

Finally, it is instructive to examine the power spectrum of the
$F^\mathrm{dm}_\mathrm{real}$ field, which is shown as a dashed curve
in Figure \ref{fig:freal_fromgasdm}.  As expected, the power spectrum
of $F^\mathrm{dm}_\mathrm{real}$ shows no evidence for a small-scale
cut-off: because the dark matter is not smoothed by pressure there is
power on arbitrarily small scales.  The power spectrum at large scales
is parallel to that for $F_\mathrm{real}$ obtained from the gas
distribution but the amplitude is lower.\footnote{The
  $F_\mathrm{real}$ and $F^\mathrm{dm}_\mathrm{real}$ fields are
  non-linear transformations of the gas and dark matter density fields
  respectively.  Expressing the power spectrum of these non-linear
  quantities in terms of the original field requires computing a
  complicated mode-coupling integral, which couples large and small
  scale power.  As such the reduced amplitude of
  $F^\mathrm{dm}_\mathrm{real}$ relative to $F_\mathrm{real}$ is
  related to the different shapes of the gas and dark matter power
  spectra.  The reduced amplitude of $F^\mathrm{dm}_\mathrm{real}$ is
  intuitive when one realizes that both $F_\mathrm{real}$ and
  $F^\mathrm{dm}_\mathrm{real}$ span the same domain from zero to
  unity and have nearly identical mean values, whereas the variance of
  $F^\mathrm{dm}_\mathrm{real}$ is comparable to that of
  $F_\mathrm{real}$ (although slightly higher because of additional
  small-scale power). The variance of a field is $\sigma^2 =
  1/2\pi\int\Delta^2(k)d\log k$, which is proportional to the area
  under the curves in Figure~\ref{fig:freal_fromgasdm}. Thus given the
  extra small-scale power in $F^\mathrm{dm}_\mathrm{real}$, in order
  for $F^\mathrm{dm}_\mathrm{real}$ and $F_\mathrm{real}$ to have
  comparable variance, the overall amplitude of
  $F^\mathrm{dm}_\mathrm{real}$ must be lower.}  This comparison
demonstrates that the cut-off in the power spectrum of the
$F_\mathrm{real}$ distribution in our simulations results from
pressure smoothing.

\subsection{Quantifying the pressure smoothing scale}

We have demonstrated that the power spectrum of the $F_\mathrm{real}$
field shows a conspicuous small-scale cut-off, which we have
interpreted as an effect of pressure smoothing. In this section we
quantify the location of this cut-off, and demonstrate that its
location behaves as expected from the classical Jeans scale argument.

\begin{figure}
  \begin{center}
    \includegraphics*[width=\columnwidth]{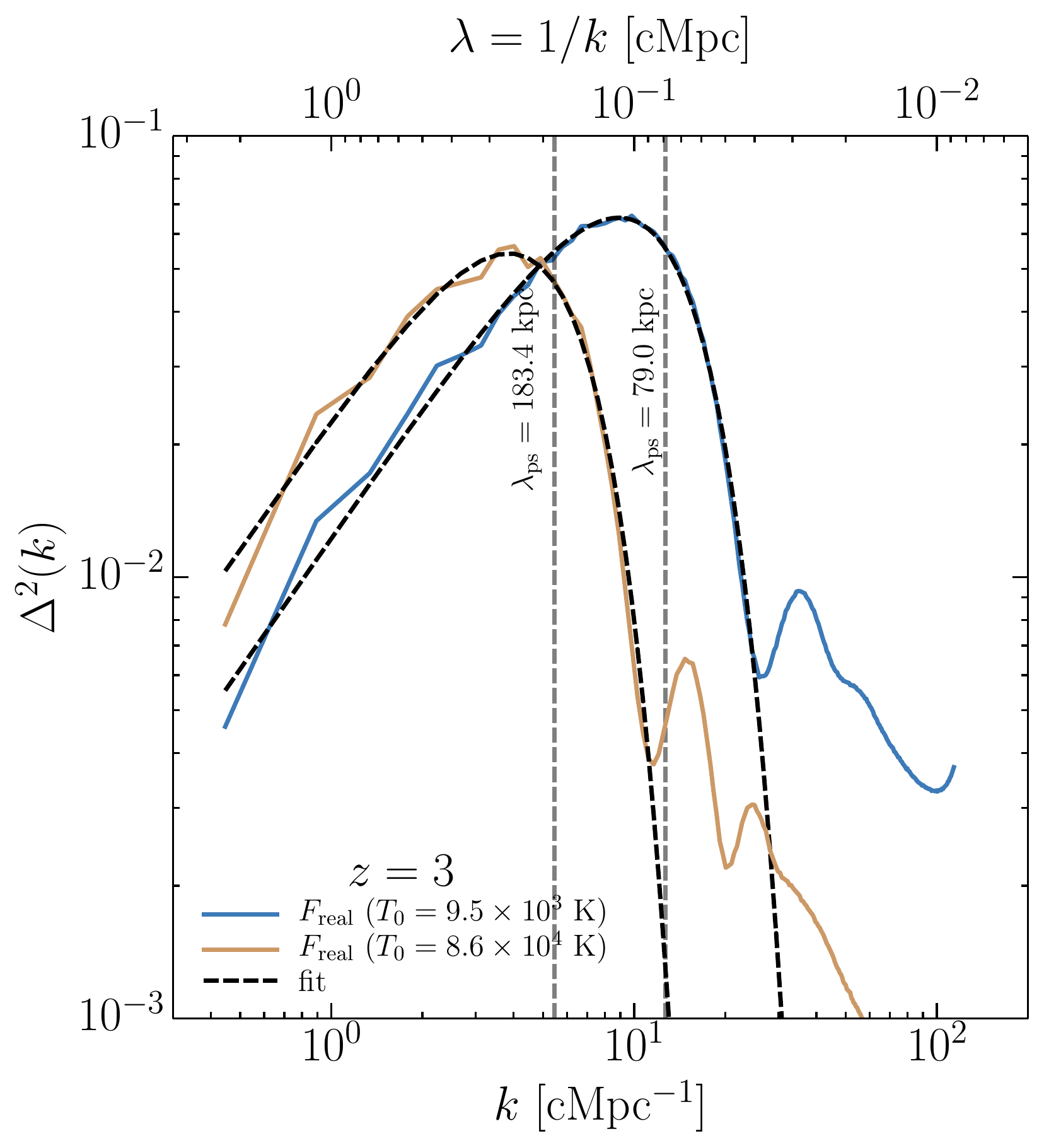}
    %  fit_freal_ps.py case 1 
  \end{center}
  \caption{Fits to $F_\mathrm{real}$ power spectra from two
    simulations with different temperatures showing that the cut-off
    in the power spectrum is Gaussian.  Blue and beige curves show the
    $F_\mathrm{real}$ power spectra from the simulations; black dashed
    curves are fits.  The effect of temperature on the cut-off is
    evident.}
  \label{fig:frealfit_2histories}
\end{figure}

Inspection of the $F_\mathrm{real}$ power spectrum in
Figure~\ref{fig:freal} motivates a simple fitting function of a power
law power spectrum cut off by a Gaussian:
\begin{equation}
  \Delta_\mathrm{F}^2(k)=Ak^n\exp\left(-\frac{k^2}{k_\mathrm{ps}^2}\right),
  \label{eqn:fitfn} 
\end{equation}
which has three parameters: $A$, $n$, and $k_\mathrm{ps}$.  Figure
\ref{fig:frealfit_2histories} shows that Equation~(\ref{eqn:fitfn}) is
an excellent fit to the power spectra of $F_\mathrm{real}$ in our
simulations.  The fitting function allows us to quantify the pressure smoothing
scale in our fiducial simulation ($T_0=9.5\times 10^3$~K) as
$\lambda_\mathrm{ps}=1/k_\mathrm{ps}=79.0$ comoving kpc at $z=3$.  In
the simulation in which the temperature at the mean density of the
IGM, $T_0=8.6\times 10^4$~K, is higher than that in the fiducial
simulation by a factor of $\sim 10$, is $\lambda_\mathrm{ps}=191.5$
comoving kpc.

\begin{figure}
  \begin{center}
    \includegraphics*[width=\columnwidth]{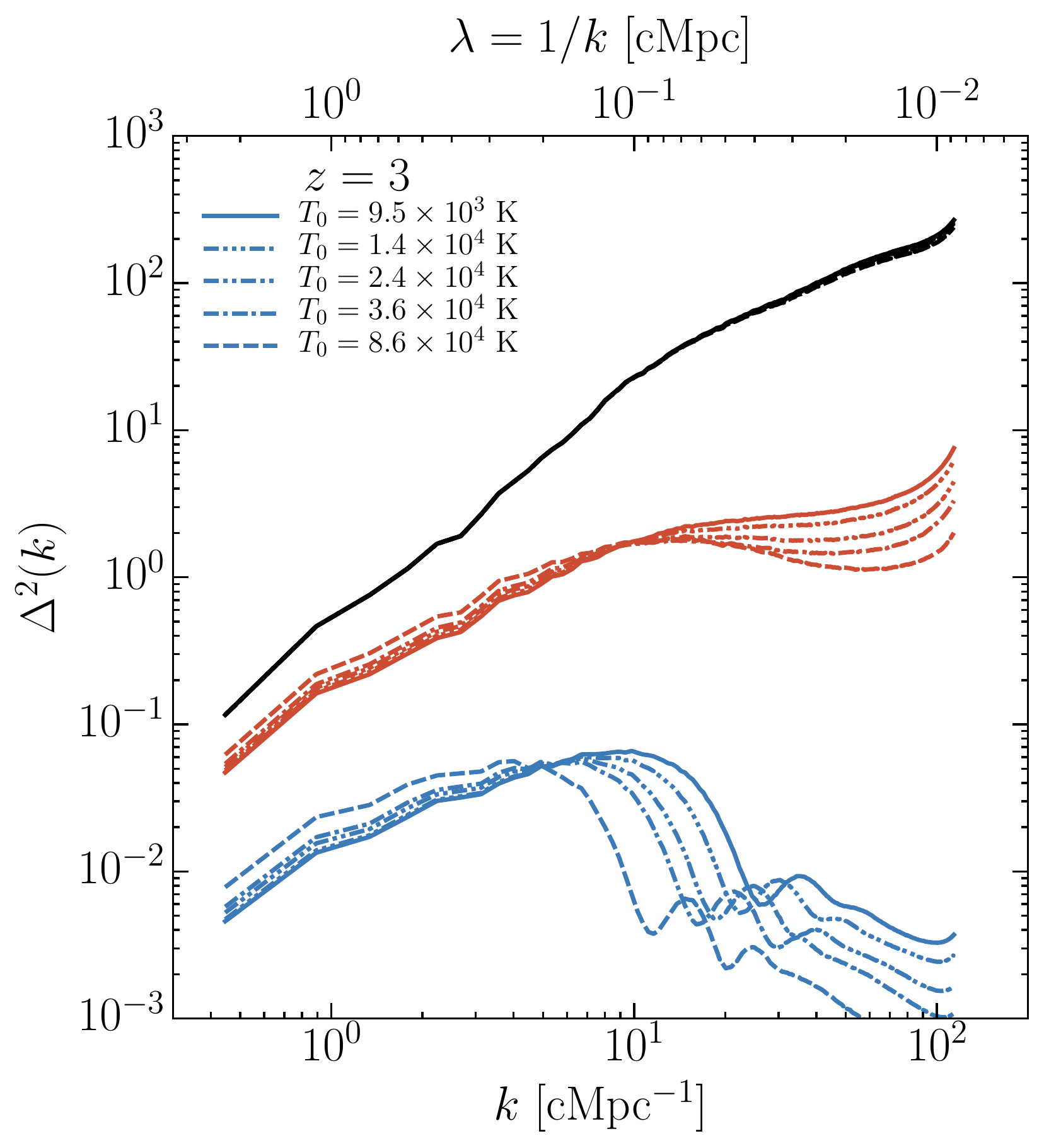}
    % ps.py case 15 
  \end{center}   
  \caption{Power spectra of the dark matter density, gas density, and
    $F_\mathrm{real}$ fields at $z=3$ in our five simulations with
    different thermal histories.  The dark matter density power
    spectra are identical for all temperatures.  The gas density power
    spectrum shows some change due to the change in temperature, but
    these small differences would change if galaxy formation were
    treated differently.  Much more significant changes are seen in
    the $F_\mathrm{real}$ power spectrum, where the cut-off scale is
    observed to progressively increase with temperature.}
  \label{fig:freal2sims}
\end{figure}

In Figure \ref{fig:freal2sims}, we compare the $F_\mathrm{real}$ power
spectra in five simulations with different temperatures ranging from
$T\sim 10^4$ to $\sim 10^5$.  As expected, the dark matter density
power spectra are essentially identical for all temperatures.  The gas
density power spectrum shows some change due to the change in
temperature, but these small differences would be masked if galaxy
formation were treated differently (see
Figure~\ref{fig:psthresh_onlydensity}), as we argued in
Section~\ref{sec:high_density_regions}.  Much more significant changes
are seen in the $F_\mathrm{real}$ power spectrum, where the cut-off
scale is observed to progressively increase with temperature.  Thus
$F_\mathrm{real}$ successfully captures a temperature-dependent
smoothing of the gas density field.

\begin{figure}
  \begin{center}
    \includegraphics*[width=\columnwidth]{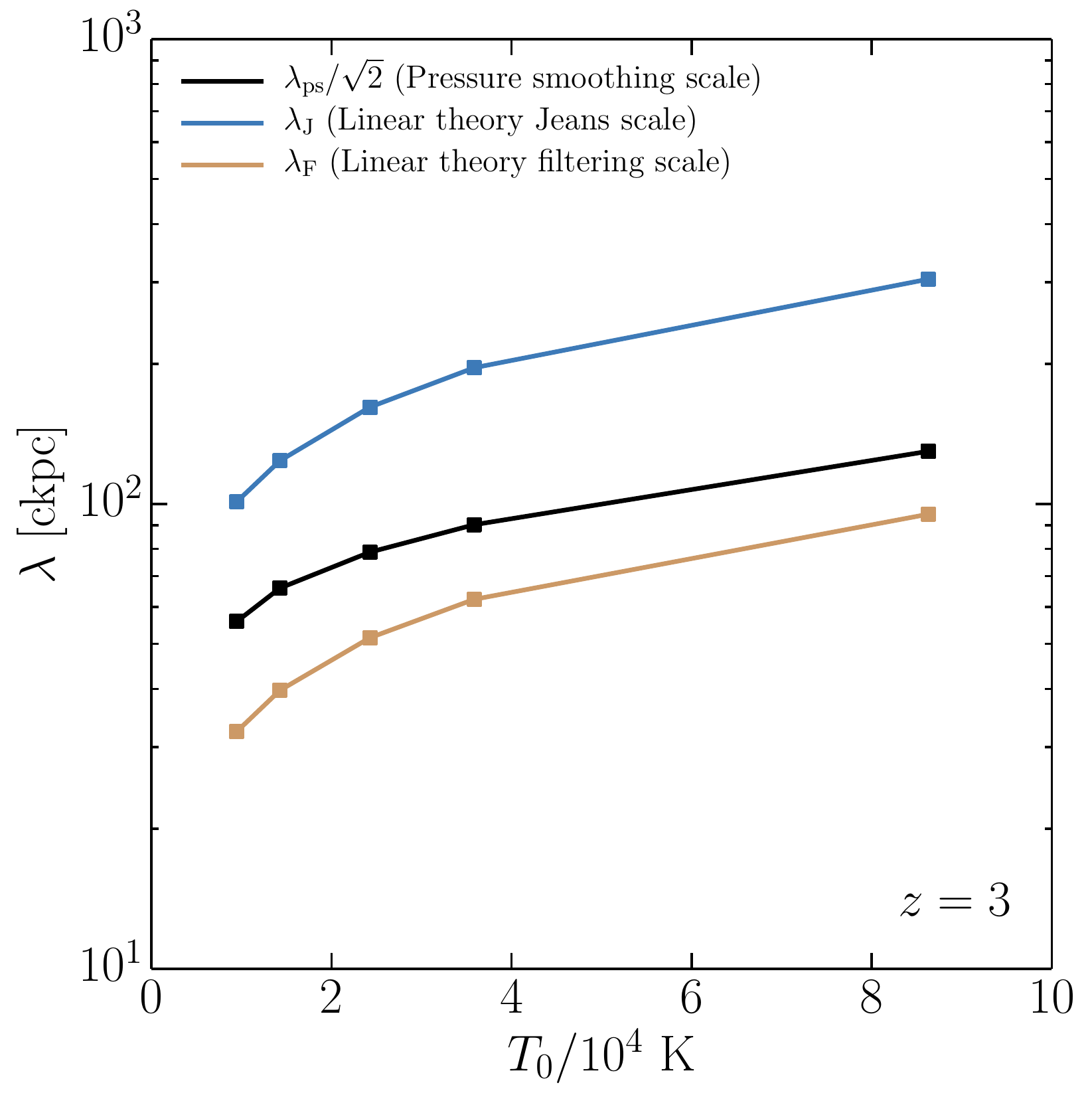}
     % freal_TempDependence.py
  \end{center}
  \caption{Temperature
  dependence of the pressure smoothing scale at $z=3$ (black curve) in
  five of our simulations with different values of $T_0$.  The
  pressure smoothing scale increases with increasing $T_0$.  The blue
  and brown curves show the temperature dependence of the linear
  theory Jeans scale, $\lambda_\mathrm{J}\propto \sqrt{T_0}$, and the
  linear theory filtering scale, $\lambda_\mathrm{F}$, defined in
  Equations~(\ref{eqn:jeans}) and (\ref{eqn:kf}), respectively.  The
  temperature dependence of the pressure smoothing scale is nearly
  identical to the $\sqrt{T_0}$ dependence of the linear theory Jeans
  and filtering scales.  (The pressure smoothing scale is defined by
  fitting the form in Equation~(\ref{eqn:fitfn}) to the flux power
  spectrum, whereas the other two scales in this figure are defined in
  terms of density contrasts.  Therefore, for consistency, the
  pressure smoothing scale has been divided by $\sqrt{2}$.)}
\label{fig:jeans_temp}
\end{figure}

The temperature dependence of the pressure smoothing scale
$\lambda_\mathrm{ps}\equiv 1/k_\mathrm{ps}$ as determined by fitting
Equation~(\ref{eqn:fitfn}) to the $F_\mathrm{real}$ power spectrum of
simulations with varying gas temperature is shown in Figure
\ref{fig:jeans_temp}. The blue and brown curves show the temperature dependence of the
linear theory Jeans scale, $\lambda_\mathrm{J}\propto \sqrt{T_0}$, and
the linear theory filtering scale, $\lambda_\mathrm{F}$, defined in
Equations~(\ref{eqn:jeans}) and (\ref{eqn:kf}), respectively.  Both of
these quantities were derived by measuring the thermal evolution of
gas in the simulations.  As expected at this post-reionization
redshift ($z=3$), the filtering scale is smaller than the Jeans scale,
by a factor of $\sim 3$ in this case \citep{1998MNRAS.296...44G,
2003ApJ...583..525G}.  We see that the temperature dependence of the
pressure smoothing scale defined by our fitting function in
Equation~(\ref{eqn:fitfn}) is nearly identical to the $\sqrt{T_0}$
dependence of the linear theory Jeans and filtering scales, which
clearly demonstrates that it is probing pressure support in the IGM.
Note that the slight deviation of our $\lambda_\mathrm{ps}$ values
from the exact $\sqrt{T}$ dependence is partly due to the fact that
simulations with different temperatures actually probe somewhat
different gas densities.  This occurs because we always rescale the UV
background of these simulations to have the same observed mean flux
value, and by changing the UV background, one changes the mapping
between flux and density, and hence the range of densities probed by
each model.  This also explains why the pressure smoothing scale is
higher than the filtering scale, as the former probes pressure
smoothing at higher densities than the mean density and therefore at
higher temperatures than $T_0$ \citep{2001ApJ...559..507S}.

This simple fitting function for $F_\mathrm{real}$ in
Equation~(\ref{eqn:fitfn}) now allows us to quantify the pressure
smoothing of the IGM, and make meaningful statements about the
pressure smoothing scale for any thermal model and reionization history.

\section{Conclusions}
\label{sec:conc}

In this paper, we presented high-resolution hydrodynamical simulations
and provided a method to characterize the pressure smoothing scale of the
IGM. We propose a new statistic, which we call the real-space flux,
$F_\mathrm{real}$, which is related to observations of correlated
Ly$\alpha$ forest absorption in close quasar pairs and is key to their
interpretation.  Our chief conclusions are:
\begin{itemize}
\item The structure of the IGM in hydrodynamical simulations is very
  different from linear theory expectations at redshifts probed by the
  Ly$\alpha$ forest. The gas density power spectrum does not exhibit
  any apparent small-scale cut-off, which is the expected signature of
  pressure smoothing in the IGM.

\item This expected cut-off in the gas power spectrum is buried under
  the contribution of dense gas in collapsed halos, which dominates
  the small-scale power.  As this dense gas is governed by the highly
  uncertain physics of galaxy formation and the circumgalactic medium,
  disentangling the high and low density regions of the baryonic field
  is crucial to reveal pressure smoothing.

\item We have introduced a new field, the real-space Ly$\alpha$ flux,
  $F_\mathrm{real}$, which naturally suppresses high density regions
  associated with galactic halos, and provides an unambiguous
  characterization of pressure smoothing in the IGM.  The power
  spectrum of $F_\mathrm{real}$ cuts off at a well-defined scale,
  which depends on gas temperature as expected for the classical Jeans
  scale, making $F_\mathrm{real}$ an effective tool for studying IGM
  pressure smoothing.

\item We provide a fitting function that accurately describes the
  power spectrum of $F_\mathrm{real}$, and defines
  the pressure smoothing scale as the small-scale cut-off of
  the $F_\mathrm{real}$ power.  This characterization of pressure
  smoothing enables one to measure the IGM pressure smoothing
  scale from any hydrodynamical simulation, and thus ask meaningful
  questions about its dependence on the thermal and reionization
  history.  Furthermore, it is directly related to observational
  measurements of correlated Ly$\alpha$ forest absorption in close
  quasar pair spectra.
\end{itemize}

Whereas to date, studies of the thermal state of the IGM have focused
largely on the amplitude $T_0$ and slope $\gamma$ of the
temperature-density relation at a given redshift $z$, the results of
this paper enable one to augment this standard description of the IGM
with a third parameter $\lambda_\mathrm{ps}$ that describes the
pressure smoothing scale, which depends on the full thermal and
reionization history. This characterization of the pressure smoothing
scale is crucial for interpreting current measurements of the thermal
state of the IGM from line-of-sight observations of the Ly$\alpha$
forest, and is also critical for interpreting measurements of the
pressure smoothing scale from quasar pair observations.

Indeed, measurements of the pressure smoothing scale using close
quasar pairs hold tremendous promise (\textcolor{linkcolor}{Rorai et
al.~2013, Rorai et al.~in prep.}). The sensitivity of the pressure
smoothing scale to the thermal and reionization history of the IGM
will be discussed in an upcoming paper (\textcolor{linkcolor}{O\~norbe
et al.~in prep.}).  We will also study the requirements for resolving
the pressure smoothing scale in hydrodynamical simulations, and the
convergence criteria for simulating a grid of IGM models with
different thermal histories (\textcolor{linkcolor}{O\~norbe et al.~in
prep.}). Measurements of the pressure smoothing scale and comparison
to these types of theoretical modeling provides a new window for
understanding the small-scale structure and thermal and reionization
history of the IGM. The $F_\mathrm{real}$ statistic introduced in this
paper provides an important new technique for relating simulations of
the low-density IGM to this new observable.

\section*{Acknowledgments}

We acknowledge helpful comments from the referee, Nick Gnedin, and
useful discussions with Martin Haehnelt, Nishikanta Khandai, Zarija
Luki\'c, Peter Nugent, Ewald Puchwein, Casey Stark, Greg Stinson, and
Martin White.  JFH acknowledges generous support from the Alexander
von Humboldt foundation in the context of the Sofja Kovalevskaja
Award. The Humboldt foundation is funded by the German Federal
Ministry for Education and Research.

\bibliography{refs}
\end{document}